\documentclass[12pt]{article}
\usepackage{bbm}
\usepackage{bbding}
\usepackage{amsmath}
\usepackage{amssymb}
\usepackage{amsfonts}
\usepackage{mathrsfs}
\usepackage{mathrsfs, amssymb}
\usepackage{booktabs}
\usepackage[perpage,symbol*]{footmisc}
\usepackage{graphicx}
\usepackage{multirow}
\usepackage[level]{datetime}
\usepackage[a4paper]{geometry}

\newcounter{theorem}
\newtheorem{theorem}{Theorem}[section]
\newcounter{lemma}

\newcounter{remark}
\newtheorem{remark}{Remark}[section]
\newcounter{example}

\newcounter{definition}

\newcounter{corollary}

\newcounter{proposition}
\newtheorem{proposition}{Proposition}[section]
\newcounter{assumption}

\newcounter{condition}

\newcounter{algorithm}

\makeatletter
\newcommand*{\indep}{%
  \mathbin{%
    \mathpalette{\@indep}{}%
  }%
}
\newcommand*{\nindep}{%
  \mathbin{
    \mathpalette{\@indep}{\not}
  }%
}
\newcommand*{\@indep}[2]{%
  \sbox0{$#1\perp\m@th$}
  \sbox2{$#1=$}
  \sbox4{$#1\vcenter{}$}
  \rlap{\copy0}
  \dimen@=\dimexpr\ht2-\ht4-.2pt\relax
  \kern\dimen@
  {#2}%
  \kern\dimen@
  \copy0 
}
\makeatother

\def\vtheta{\mbox{\boldmath$\theta$}}

\def\vlambda{\mbox{\boldmath$\lambda$}}

\def\b{\mbox{\boldmath$b$}}
\def\s{\mbox{\boldmath$s$}}
\def\u{\mbox{\boldmath$u$}}
\def\U{\mbox{\boldmath$U$}}

\def\x{\mbox{\boldmath$x$}}
\def\y{\mbox{\boldmath$y$}}

\def\var{\mbox{var}}

\def\prow{\stackrel{\textstyle p}{\longrightarrow}}
\def\asrow{\stackrel{a.s.}{\longrightarrow}}

\begin{document}
\baselineskip 18pt
\begin{center}
{\large\bf Semiparametric mean and variance joint models with Laplace link functions for count time series}
\end{center}

\begin{center}
{ \textbf{\small{ Tianqing Liu}}$^*$}\\
{ \small{Center for Applied Statistical Research and School of Mathematics, Jilin University, China}}\\
{ \small{\texttt{\textsc{tianqingliu@gmail.com}}}}

{ \textbf{\small{ Xiaohui Yuan}}}\\
{ \small{School of Mathematics and Statistics, Changchun University of Technology, China}}\\
{ \small{\texttt{\textsc{yuanxh@ccut.edu.cn}}}}
\end{center}
\begin{center}
This version: \usdate\today
\end{center}

\footnotetext{$^*$Corresponding author, $^\dag$ equal authors contribution.}

\begin{abstract}
{\small Count time series data are frequently analyzed by modeling their conditional means and the conditional variance is often considered to be a deterministic function of the corresponding conditional mean and is not typically modeled independently.  We propose a semiparametric mean and variance joint model, called random rounded count-valued generalized autoregressive conditional heteroskedastic (RRC-GARCH) model, to address this limitation.  The RRC-GARCH model and its variations allow for the joint modeling of both the conditional mean and variance and offer a flexible framework for capturing various mean-variance structures (MVSs). One main feature of this model is its ability to accommodate negative values for regression coefficients and autocorrelation functions. The autocorrelation structure of the RRC-GARCH model using the proposed Laplace link functions with nonnegative regression coefficients is the same as that of an autoregressive moving-average (ARMA) process. For the new model, the stationarity and ergodicity are established and the consistency and asymptotic normality of the conditional least squares estimator are proved. Model selection criteria are proposed to evaluate the RRC-GARCH models. The performance of the RRC-GARCH model is assessed through analyses of both simulated and real data sets. The results indicate that the model can effectively capture the MVS of count time series data and generate accurate forecast means and variances.
\paragraph{\small Keywords:} Conditional mean, Conditional
variance, Count time series, Integer-valued time series, Random rounding
operator}
\end{abstract}

\section {Introduction} \label{sec1} \setcounter {equation}{0}
\def\theequation{\thesection.\arabic{equation}}

Gaussian models have been widely used for analyzing real-valued time series data. Because these models are completely characterized by their first two moments and
and the conditional mean and variance can be modeled separately. However, Gaussian models often poorly described discrete-valued series. Various models (Kedem \& Fokianos, 2002; Cameron \& Trivedi, 2013; Wei${\ss}$, 2018) have been proposed for modeling discrete-valued data, taking into account their specific characteristics and distributions.

Integer-valued time series models have garnered increasing interest in recent years due to their applicability in various domains such as finance, economics, and telecommunications. These models are defined within the set of all integers, i.e., $\mathbb{Z} = \{...,-2,-1,0,1,2,...\}$.
Liu \& Yuan (2013) showed a mean-variance structure (MVS) for all integer-valued data. Let $\{X_t,\ t\in\mathbb{Z}\}$ be an integer-valued time series and
$\mathcal {F}_{t-1}$ be the $\sigma$-field generated by $\{X_{t-1},X_{t-2},\cdots,\}$. The MVS is given by
\begin{eqnarray}\label{rrarmameanvar}
\var(X_t|\mathcal{F}_{t-1})=R(E(X_t|\mathcal{F}_{t-1}))+h_t,
\end{eqnarray}
where $h_t$ is a non-negative $\mathcal {F}_{t-1}$-measurable function and for $c\in \mathbb{R}$,
\begin{eqnarray}\label{Rfun}
R(c)=(\Delta(c)+1-c)(c-\Delta(c))\ \text{with}\ \Delta(c)=\max\{z\in \mathbb{Z}: z\leq c\}.
\end{eqnarray}
Thus, the inequality $\var(X_t|\mathcal{F}_{t-1})\geq R(E(X_t|\mathcal{F}_{t-1}))$ holds for all integer-valued time series. Integer-valued time series have other traits frequently observed in practice, such as overdispersion or underdispersion, excess
of zeros or ones, asymmetry or symmetry of marginal distribution, and persistence (Wei${\ss}$, 2018). Numerous integer-valued models have been introduced so far to highlight the above nature of integer-valued time series. See Scotto et al. (2015), Karlis \& Mamode (2023) and Li et al. (2024) for a comprehensive review for integer-valued models.

Count time series take values in $\mathbb{Z}_+ = \{0,1,2,...\}$ and are important cases of integer-valued time series. Count time series arise in numerous applied scientific areas and usually count of some event in time and/or space. Originally, count series were often described via thinning operator (McKenzie, 1985, 2003) or regression type models (Davis et al., 2000, 2003; Davis \& Wu, 2009; Kedem \& Fokianos, 2002). As the field developed, various approaches for modeling count series emerged, such as the multiplicative error models (Heinen, 2003;  Aknouche \& Scotto, 2024; Wei${\ss}$ \& Zhu,  2024), the copula-based models (Jia et al., 2023; Kong \& Lund, 2023), and the discrete mixed models (Gorgi, 2020; Chen et al., 2022; Maya et al., 2022). However, no single class of models dominates the count time series landscape and the field developed without a unifying statistical model, estimation method or inference theory.

All the above models are conditional mean-based models and the conditional variance is a deterministic function of the conditional mean. However, specifying the conditional MVS in a real data analysis can be challenging. While mean-based models are often easier to conceptualize and implement, accurately capturing the conditional variance adds another layer of complexity. In count time series analysis, the conditional MVS plays a central role in inference processes. However, due to the inherent complexities of real-world data, precisely defining this structure can be elusive. In practice, a misspecified MVS can indeed lead to significant inference errors. To deal with this problem, we first give two important properties of count random variables in the following proposition.

\begin{proposition}\label{countprop}
(Count properties) Let $X$ be a count random variable taking values in $\mathbb{Z}_+ $ such that the mean $\mu=E(X)$ and variance $\sigma^2(\mu)=E(X-\mu)^2$ exist. By Markov's inequality, $\mathbb{P}(X=0|\mu)=1-\mathbb{P}(X>\epsilon_0|\mu)\geq 1-\min(1,\mu/\epsilon_0)$ and $\mathbb{P}(X\in \{0,1\}|\mu)=1-\mathbb{P}(X>\epsilon_1|\mu)\geq 1-\min(1,\mu/\epsilon_1)$ for any $0<\epsilon_0<1$ and $1<\epsilon_1<2$. Thus, $(\mathbbm{a})$: $\lim_{\mu\rightarrow0^+}\mathbb{P}(X=0|\mu)=\lim_{\mu\rightarrow0^+}\mathbb{P}(X\in \{0,1\}|\mu)=1$ and $(\mathbbm{b})$: $\lim_{\mu\rightarrow0^+}\sigma^2(\mu)=0$.
\end{proposition}

Obviously, count property $(\mathbbm{a})$ can explain excess of zeros or ones of observations of $X$ with a small $\mu>0$ and property $(\mathbbm{b})$ describes the MVSs of count random variables, which implies that, for a count random variable, its variance $\sigma^2(\mu)$ is controlled by its mean $\mu$, when $\mu$ is small. In practice, one may apply the integer-valued models to analyze the count data. However, an ideal count model should not only preserve nonnegative characteristics but also adhere to properties $(\mathbbm{a})$ and $(\mathbbm{b})$, thereby potentially offering superior fit or prediction for count time series data.

In the spirit of Box (1976), all models are wrong, but some are useful. A useful integer-valued or count-valued time series model should (a) be capable of capturing and generating a wide range of patterns within the time series data, (b) offer flexibility in the MVSs it can accommodate, (c) demonstrate its good predictive power when applied to real-world datasets. To account for these points, we propose a semiparametric count time series model, called RRC-GARCH. The main advantages of the new model can be summarized as follows.
\begin{itemize}
\item[(a)]  The RRC-GARCH model offers a novel semiparametric framework to
jointly model the conditional mean and variance for count time series.
\item[(b)] Both the RRC-GARCH$(p_1,p_2)$ process with nonnegative regression coefficients and an ARMA$(p,p_2)$ process exhibit the same autocorrelation structure, where $p=\max\{p_1,p_2\}$.
\item[(c)] The RRC-GARCH model is more flexible than the existing models as its variants can provide various MVSs.
\item[(d)] The RRC-GARCH model allows negative values both for its regression coefficients and
autocorrelation function.
\end{itemize}

The rest of the paper is organized as follows. We first propose the RRC-GARCH model in Section 2. Then, in Section 3, we give theoretical
results about the stationarity, ergodicity and autocorrelation structure of the
RRC-GARCH$(p_1,p_2)$ process. Next, in Section 4, we prove the consistency
and asymptotic normality of the conditional least squares estimator.
In Section 5, we discuss the problem of model selection for the
RRC-GARCH models. Simulations are presented in Section 6. In Section
7, applications of the model to two real data sets are discussed.
Finally, we conclude the paper in Section 8. The proofs of all
forthcoming results are postponed to the Section S1 of the Supplementary Material.

\section{Background and the RRC-GARCH model} \setcounter {equation}{0}
\def\theequation{\thesection.\arabic{equation}}

In the context of integer-valued time series, Liu \& Yuan (2013) first proposed a semiparametric GARCH-type model to separately model the conditional mean and variance. In the following, we first give a simple review of the RRIN-GARCH models (Liu \& Yuan, 2013; Li et al., 2024) and then propose our count model.

\subsection{RRIN-GARCH model}

Liu \& Yuan (2013) proposed two random rounding operators. Let
\begin{eqnarray*}
B(x)=\frac{[\Delta(x^{1/2})+1]^2-x}{[\Delta(x^{1/2})+1]^2-[\Delta(x^{1/2})]^2},
\ x\in\mathbb{R}_+=\{x\in\mathbb{R}:x\geq0\},
\end{eqnarray*}
with $\Delta(\cdot)$ given in (\ref{Rfun}). The first-order and second-order random rounding operators are
defined respectively as
\begin{eqnarray}
\odot_1(x,U)&=&\Delta(x)+{1}(U\geq 1+\Delta(x)-x),\ x\in
\mathbb{R},\label{round1}\ \ \text{and}\\
\odot_2(x,U)&=&\Delta(x^{1/2})+{1}(U\geq B(x)),\
x\in\mathbb{R}_+,\label{round2}
\end{eqnarray}
where ${1}(A)$ is the indicator function of $A$ and $U$ is a uniform random variable defined on the interval
$[0,1]$. A RRIN-GARCH model is given as
\begin{eqnarray}\label{rrarma}
\left\{ \begin{aligned}
         &X_t=\odot_1(\mu_t,U_{1t}) +\varepsilon_t,\  \ t\in \mathbb{Z},\\
         &\varepsilon_t=\odot_2(h_{t},U_{2t})\varsigma_t,\  \ t\in \mathbb{Z},
                          \end{aligned} \right.
\end{eqnarray}
where the variables $U_{1t}$, $U_{2t}$ and $\varsigma_t$, $(t\in\mathbb{Z})$
are mutually independent; $(U_{1t})$ and $(U_{2t})$ are two sequences of independent and identically distributed (i.i.d.) uniform random
variables defined on the interval $[0, 1]$; and $(\varsigma_t)$ is a sequence of i.i.d. integer-valued random variables with range $\mathbb{Z}$, mean 0 and finite variance 1. Different assumptions on $\mu_t$ and $h_t$ lead to different integer-valued time series models (Liu \& Yuan, 2013; Li et al., 2024). The MVS of RRIN-GARCH model is given in (\ref{rrarmameanvar}) with $E(X_t|\mathcal {F}_{t-1})=\mu_t$. However, the RRIN-GARCH model in (\ref{rrarma}) does not possess the nonnegative property and count properties in Proposition \ref{countprop}.

\subsection{RRC-GARCH model}

Let $(X_t)$, $t\in\mathbb{Z}$ be a count process, where $X_t$'s are (unbounded) count random variables with range $\mathbb{Z}_+$. For $x\in\mathbb{R}$, define $L_0(x)=\max(x,0)=x^+$ as the ReLU activation function. Assume that $\mu_t=E(X_t|\mathcal {F}_{t-1})$ satisfies the following iterative equation
\begin{eqnarray}
\mu_t&=&M\biggr(c+\sum_{i=1}^{p_1}\phi_iX_{t-i}+\sum_{j=1}^{p_2}\psi_j\mu_{t-j}\biggr),\  \ t\in\mathbb{Z},\label{rinar0}
\end{eqnarray}
where $p_1\geq1$, $p_2\geq0$, $\phi_{p_1},\psi_{p_2}\neq0$, $M(\cdot)$ is a specified link function satisfies that there exists a constant $s_0\in\mathbb{R}_+$ such that $0\leq M(u)\leq s_0+L_0(u)$, for $u\in\mathbb{R}$. Then, based on (\ref{rinar0}), we propose the following count time series
model:
\begin{eqnarray}\label{rinar1}
\left\{ \begin{aligned}
         &X_t=\odot_1(\mu_t,U_{1t}) +\varepsilon_t,\  \ t\in \mathbb{Z},\\
         &\varepsilon_t=\odot_2(\kappa_{t},U_{2t})(\zeta_t-1),\\
         &\kappa_t=\{\odot_1(\mu_t,U_{1t})\}^{2\tau},
                          \end{aligned} \right.
\end{eqnarray}
where $0<\tau\leq1$, the first-order and second-order random rounding operators
$\odot_1$ and $\odot_2$ are defined respectively in (\ref{round1})
and (\ref{round2}); $(\zeta_t)$ is a sequence of i.i.d.
count random variables with range $\mathbb{Z}_+$, mean 1 and finite variance $\sigma_\zeta^2$;
$U_{kt}$ with $k\in\{1,2\}$ and $t\in\mathbb{Z}$ are i.i.d. uniform random
variables defined on the interval $[0, 1]$; and the variables $U_{1t}$, $U_{2t}$ and $\zeta_t$, $(t\in\mathbb{Z})$
are mutually independent. We call this model RRC-GARCH$(p_1,p_2)$. In the RRC-GARCH$(p_1,p_2)$ model (\ref{rinar1}), the unknown parameter vectors $\vtheta=(c,\phi_1,\cdots,\phi_{p_1},\psi_1,\cdots,\psi_{p_2})^\textsf{T}$ and $\vlambda=(\tau,\sigma_\zeta^2)^\textsf{T}$ determine the MVS of $X_t$. However, the distribution of $\zeta_t$ is not specified and remains nonparametric. In the following proposition, we present our findings
on the basic properties of the RRC-GARCH models.

\begin{proposition}\label{ranround2}
The following equalities hold:
\begin{eqnarray}
&&(a)\ \mathbb{P}(X_t\in\mathbb{Z}_+)=1,\ \  (b)\ E(X_t|\mathcal{F}_{t-1})=\mu_t,\ \ \text{and}\nonumber\\
&&(c)\ \var(X_t|\mathcal{F}_{t-1})=R(\mu_t)+V_\tau(\mu_t) \sigma_\zeta^2,\label{const0}
\end{eqnarray}
where the function $R(\cdot)$ is defined in (\ref{Rfun}) and
\begin{eqnarray}\label{vtaufun}
V_\tau(\mu_t)=\{\Delta(\mu_t)\}^{2\tau}\{1+\Delta(\mu_t)-\mu_t\}+\{1+\Delta(\mu_t)\}^{2\tau}\{\mu_t-\Delta(\mu_t)\}.
\end{eqnarray}
Moreover, for fixed $\tau>0$, $V_\tau(x)$ is a continuous increasing function of $x$ on $\mathbb{R}_+$; for any fixed $x\geq0$, $V_\tau(x)$ is a non-decreasing function of $\tau$ on $(0,+\infty)$; $V_{0.5}(x)\equiv x$ and $V_1(x)\equiv R(x)+x^2$ for $x\geq0$; $V_\tau(x)\equiv x$ for $x\in[0,1]$ and $\tau>0$; and $V_\tau(x)\equiv x^{2\tau}$ for $x\in \mathbb{Z}_+$ and $\tau>0$. Overall, $V_\tau(x)$ is a piecewise linear approximation of $x^{2\tau}$. In Figure 1, we present the plots of the functions $V_\tau(\cdot)$ with different values of $\tau$.
\end{proposition}

\begin{remark}\label{softplus}
A convenient choice for the link function is the ReLU activation function, i.e., $M(u)=L_0(u)$. Other non-linear link specifications are also applicable for $\mu_t$ in (\ref{rinar0}), e.g., by adapting the softplus structure (Dugas et al., 2000; Mei $\&$ Eisner,2017; Wei${\ss}$ et al., 2022). The softplus function is given by $S_\sigma(u)=\sigma\log\{1+\exp(u/\sigma)\}$, where $u\in\mathbb{R}$ and $\sigma>0$. In this paper, we propose the following Laplace link function:
\begin{eqnarray}
&&L_\sigma(u)=-\sigma\log\{1-F(u/\sigma)\}\label{Laplace}\\
&=&-\sigma\log\{1-0.5\exp(u/\sigma)\}{1}(u\leq0)+\{\sigma\log(2)+u\}{1}(u>0),\ u\in\mathbb{R},\ \sigma>0,\nonumber
\end{eqnarray}
where $F(u)=0.5\exp(u){1}(u\leq0)+\{1-0.5\exp(-u)\}{1}(u>0)$ is the standard Laplace cumulative distribution function (CDF) and  $\sigma = 1$ is the default choice. It is easy to verify that $L_0(u)\leq S_\sigma(u)\leq \sigma\log(2)+L_0(u)$ and $L_0(u)\leq L_\sigma(u)\leq \sigma\log(2)+L_0(u)$, where $S_\sigma(u)= \sigma\log(2)+L_0(u)$ for $u=0$ and $L_\sigma(u)= \sigma\log(2)+L_0(u)$ for $u\geq0$. Thus, $\lim_{\sigma\rightarrow0^+}L_\sigma(u)=\lim_{\sigma\rightarrow0^+}S_\sigma(u)=L(u)$. In contrast to the softplus function, the Laplace link function $L_\sigma(u)$ with $\sigma>0$ is linear for all $u \geq 0$. Moreover, $L_\sigma(\cdot)$ is a truly positive and  continuously differentiable function on whole $\mathbb{R}$, and it holds that $P_\sigma(u)=:\partial L_\sigma(u)/\partial u=f(u/\sigma)/\{1-F(u/\sigma)\}=\frac{0.5\exp(u/\sigma)}{1-0.5\exp(u/\sigma)}{1}(u\leq0)+{1}(u>0)$, where $f(u)=0.5\exp(-|u|)$ is the standard Laplace density function. In Figures 2-3, we present the plots of the functions $L_\sigma(u)$ and $P_\sigma(u)$ with different values of $\sigma$. We stress that the the formula in (\ref{Laplace}) can generate a class of distribution link functions by setting $F(\cdot)$ as a CDF such that $F(x)<1$ for $x\in\mathbb{R}$. For example, if we set $F(\cdot)$ as the Logistic CDF, we obtain the softplus link function $S_\sigma(u)$.
\end{remark}

\begin{remark}\label{meanexp}
For $M(u)=L_\sigma(u)$ $(\sigma\geq0)$, we consider two parameter spaces: $\Theta_0=\{\vtheta: c\in\mathbb{R}, \sum_{i=1}^{p_1}|\phi_i|+\sum_{j=1}^{p_2}|\psi_j|<1\}$ and $\Theta_1=\{\vtheta: c>-\sigma\log(2), \phi_1, ..., \phi_{p_1}, \psi_1, ..., \psi_{p_2}\geq0, \sum_{i=1}^{p_1}\phi_i+\sum_{j=1}^{p_2}\psi_j<1\}$. Obvioulsy, $\Theta_1\subset \Theta_0$.
If $\vtheta\in\Theta_1$, then $\mu_t=L_\sigma(c+\sum_{i=1}^{p_1}\phi_iX_{t-i}+\sum_{j=1}^{p_2}\psi_j\mu_{t-j})\equiv \sigma\log(2)+c+\sum_{i=1}^{p_1}\phi_iX_{t-i}+\sum_{j=1}^{p_2}\psi_j\mu_{t-j}\geq0$. In this case, $\mu_t$ in (\ref{rinar0}) reduces to a linear function of $(X_{t-1},X_{t-2},\cdots)$. For $M(u)=S_\sigma(u)$ $(\sigma>0)$, we consider the parameter space $\Theta_0$. For all $\vtheta\in\Theta_0$, then $\mu_t=S_\sigma(c+\sum_{i=1}^{p_1}\phi_iX_{t-i}+\sum_{j=1}^{p_2}\psi_j\mu_{t-j})$ is always a nonlinear function of $(X_{t-1},X_{t-2},\cdots)$.
\end{remark}

\begin{remark}\label{identification}
The RRC-GARCH$(p_1,p_2)$ model involves the parameter vectors $\vtheta$ and $\vlambda$. If we take the conditional mean function $M(\cdot)$ as the softplus or Laplace link function, then the parameter vector $\vtheta$ is identifiable if the conditional distribution of $X_t$ is not degenerate. Because $S_\sigma(\cdot)$ and $L_\sigma(\cdot)$ are strictly monotone increasing for $\sigma>0$. The identification of $\vlambda$ may be difficult. For example, if $\mu_t$ is small enough such that $\mathbb{P}(0\leq\mu_t\leq1)=1$, then $V_\tau(\mu_t)\equiv \mu_t$ and $\tau$ can not be identifiable. However, $\sigma_\zeta^2$ and $\var(X_t|\mathcal{F}_{t-1})=R(\mu_t)+\mu_t \sigma_\zeta^2$ are always identifiable. We stress that the conditional mean and variance functions are our main interests and are identifiable if the conditional distribution of $X_t$ is not degenerate. Moreover, if $\mathbb{P}(\mu_t>1)>0$ and the conditional distribution of $X_t$ is not degenerate, then $\tau$ is identifiable.
\end{remark}

\begin{remark}\label{Extend}
(Extended RRC-GARCH) Let $\kappa_t=\{\odot_1(\mu_t,U_{1t})\}^{2\tau}\nu_t$ in (\ref{rinar1}), where $\nu_t\in[0,1]$ is a $\mathcal {F}_{t-1}$-measurable function, e.g., $\nu_t= R(\mu_t)$ or $\nu_t=\tau$. Then, we obtain the extended MVS:
\begin{eqnarray}\label{const}
\var(X_t|\mathcal{F}_{t-1})=R(\mu_t)+\nu_t V_\tau(\mu_t)\sigma_\zeta^2,\ \ 0<\tau\leq1.
\end{eqnarray}
\end{remark}

\begin{remark}\label{Power}
(Power RRC-GARCH) Using Jesen's inequality, we have, $$V_{\tau}(\mu_t)=E[\{\odot_1(\mu_t,U_{1t})\}^{2\tau}|\mathcal{F}_{t-1}]\geq (E[\{\odot_1(\mu_t,U_{1t})\}|\mathcal{F}_{t-1}])^{2\tau}=\mu_t^{2\tau},$$ for $1\leq2\tau\leq2$; and
$$V_{\tau}(\mu_t)=E[\{\odot_1(\mu_t,U_{1t})\}^{2\tau}|\mathcal{F}_{t-1}]\leq (E[\{\odot_1(\mu_t,U_{1t})\}|\mathcal{F}_{t-1}])^{2\tau}=\mu_t^{2\tau},$$ for $0<2\tau<1$.
Let $\kappa_t=\min\{\frac{\mu_t^{2\tau}}{V_{\tau}(\mu_t)},1\}\{\odot_1(\mu_t,U_{1t})\}^{2\tau}$ in (\ref{rinar1}), we obtain the power MVS:
\begin{eqnarray}\label{const1}
\var(X_t|\mathcal{F}_{t-1})=R(\mu_t)+\min\{\mu_t^{2\tau},V_{\tau}(\mu_t)\}\sigma_\zeta^2,\ \ 0<\tau\leq1.
\end{eqnarray}
\end{remark}
\begin{remark}\label{Mixture}
(Mixture RRC-GARCH) In (\ref{rinar1}), let
\begin{eqnarray*}
\kappa_t=\frac{r}{1+r}\odot_1(\mu_t,U_{1t})+\frac{1}{1+r}\min\biggr\{\frac{\mu_t^{2\tau}}{V_{\tau}(\mu_t)},1\biggr\}\{\odot_1(\mu_t,U_{1t})\}^{2\tau},
\end{eqnarray*}
where $0<\tau\leq1$ and $r\in\mathbb{Z}_+$. Then, we obtain the mixture MVS:
\begin{eqnarray}\label{const2}
\var(X_t|\mathcal{F}_{t-1})=R(\mu_t)+\biggr[\frac{r}{1+r}\mu_t+\frac{1}{1+r}\min\{\mu_t^{2\tau},V_{\tau}(\mu_t)\}\biggr]\sigma_\zeta^2.
\end{eqnarray}
\end{remark}
\begin{remark}\label{leadingterm}
Similar to the proof of proposition \ref{ranround2}, it is easy to verify that $\mathbb{P}(X_t\in\mathbb{Z}_+)=1$ for the extended RRC-GARCH, power RRC-GARCH and mixture RRC-GARCH models. Moreover, Wei${\ss}$ $\&$ Zhu  (2024) proposed a conditional-mean multiplicative error model (CMEM) with binomial multiplicative operator, whose MVS is $\var(X_t|\mathcal{F}_{t-1})=R(\mu_t)+\mu_t^2\sigma_\zeta^2$, where $\mu_t=E(X_t|\mathcal{F}_{t-1})$ and $\sigma_\zeta^2$ is the variance of the multiplicative error. Obviously, the power MVS (\ref{const1}) with $\tau=1$ and the mixture MVS (\ref{const2}) with $r=0$ and $\tau=1$ nest the MVS of the CMEM with binomial multiplicative operator as a special case. If $\sigma_\zeta^2\rightarrow0^+$ and $0<\mu_t<1$, we have $\var(X_t|\mathcal{F}_{t-1})=R(\mu_t)=\mu_t(1-\mu_t)$, which is the MVS of a Bernoulli distribution with mean $\mu_t$. By the definition of $R(\cdot)$, we get $0\leq R(\mu_t)\leq \min\{\mu_t-\Delta(\mu_t),\Delta(\mu_t)+1-\mu_t,1/4\}$. If $\sigma_\zeta^2$ or $\mu_t$ is large enough, the contribution of $R(\mu_t)$ to $\var(X_t|\mathcal{F}_{t-1})$ is negligible in (\ref{const1}) and (\ref{const2}). In such a situation, the conditional variance in (\ref{const2}) is an approximation of $\bigr[\frac{r}{1+r}\mu_t+\frac{1}{1+r}\min\{\mu_t^{2\tau},V_{\tau}(\mu_t)\}\bigr]\sigma_\zeta^2$, which contains the MVSs of Poisson $(\tau=0.5, \sigma_\zeta^2=1, r\in \mathbb{Z}_+)$ and negative binomial distributions $(\tau=1, \sigma_\zeta^2=1+r^{-1}, r\in \mathbb{Z}_+\setminus\{0\})$.
\end{remark}
\begin{remark}\label{posscount}
By the definitions of $\odot_1(\mu,U)$, $\odot_2(\mu,U)$, $R(\mu)$ and $V_\tau(\mu)$, it is easy to verify that $\mathbb{P}\{\lim_{\mu\rightarrow0^+}\odot_1(\mu,U)=0\}=\mathbb{P}\{\lim_{\kappa\rightarrow0^+}\odot_2(\kappa,U)=0\}=1$ and $\lim_{\mu\rightarrow0^+}R(\mu)=\lim_{\mu\rightarrow0^+}V_\tau(\mu)=0$, for $\tau>0$. It follows that, for the RRC-GARCH, extended RRC-GARCH, power RRC-GARCH and mixture RRC-GARCH models: (1) $\mathbb{P}(X_t=0|\mathcal {F}_{t-1})=\mathbb{P}(X_t=0|\mu_{t})\rightarrow1$ and $\mathbb{P}(X_t\in\{0,1\}|\mathcal {F}_{t-1})=\mathbb{P}(X_t\in\{0,1\}|\mu_{t})\rightarrow1$ as $\mu_{t}\rightarrow0$; (2) their conditional variances in (\ref{const0}), (\ref{const}), (\ref{const1}) and (\ref{const2}) satisfy that $\var(X_t|\mathcal{F}_{t-1})=\var(X_t|\mu_{t})\rightarrow0$ as $\mu_{t}\rightarrow0$. In summary, the RRC-GARCH, extended RRC-GARCH, power RRC-GARCH and mixture RRC-GARCH models possess nonnegative property and count properties in Proposition \ref{countprop}.
\end{remark}

\begin{remark}\label{unify}
For the RRC-GARCH, extended RRC-GARCH, power RRC-GARCH and mixture RRC-GARCH models, their MVSs have a unified expression
\begin{eqnarray}\label{unifymvs}
\var(X_t|\mathcal{F}_{t-1})=R(\mu_t)+D_\tau(\mu_t)\sigma_\zeta^2,
\end{eqnarray}
where $D_\tau(\mu_t)=V_\tau(\mu_t)$ for the RRC-GARCH, $D_\tau(\mu_t)=V_\tau(\mu_t)\nu_t$ for the extended RRC-GARCH, $D_\tau(\mu_t)=\min\{\mu_t^{2\tau},V_{\tau}(\mu_t)\}$ for the power RRC-GARCH, and $D_\tau(\mu_t)=\frac{r}{1+r}\mu_t+\frac{1}{1+r}\min\{\mu_t^{2\tau},V_{\tau}(\mu_t)\}$ for the mixture RRC-GARCH. Here, we treat $\nu_t$ and $r$ as known terms. In Figures S1-S2 of the Supplementary Material, we present the plots of the functions $D_\tau(u)$ for the power RRC-GARCH and mixture RRC-GARCH (r=1) with different values of $\tau$.
\end{remark}
\section{Ergodicity, stationarity, autocorrelation structure and prediction}
\setcounter{equation}{0}\def\theequation{\thesection.\arabic{equation}}

In this section, we study the stationary conditions, autocorrelation
structure and prediction of the RRC-GARCH$(p_1,p_2)$ process.

\subsection{Ergodicity and stationarity}

Define
\begin{eqnarray}\label{eta}
\eta_t=\odot_1(\mu_t,U_{1t})-\mu_t,\ \ t\in
\mathbb{Z}.
\end{eqnarray}
Then, the study of the RRC-GARCH$(p_1,p_2)$ process can be carried out
through the following vectorized process
\begin{eqnarray}\label{vinarp}
\U_t=\left(
   \begin{array}{c}
X_t\\
\mu_{t}\\
\vdots\\
X_{t-p+1}\\
\mu_{t-p+1}
\end{array}\right)=\left(
   \begin{array}{c}
\mu_t+\eta_t+\varepsilon_t\\
\mu_{t}\\
\vdots\\
X_{t-p+1}\\
\mu_{t-p+1}
\end{array}\right),
\end{eqnarray}
where $p=\max(p_1,p_2)$. The process $(\U_t)$ forms a homogeneous Markov chain with state
space $\mathbb{E} =(\mathbb{Z}_+\times \mathbb{R}_+)^{p}$. For
$\x=(x_1,\cdots,x_{2p})^\textsf{T}\in\mathbb{E}$ and
$\y=(y_1,\cdots,y_{2p})^\textsf{T}\in\mathbb{E}$, the transition
probability function from $\x$ to $\y$ is given by
\begin{eqnarray*}
&&\pi(\x,\y)\\
&=&\mathbb{P}[y_1=\odot_1(\mu_1,U_{11}) +\odot_2(\kappa_{1},U_{21})(\zeta_1-1)]{1}(y_2=\mu_1,y_{3}=x_{1}\cdots,y_{2p}=x_{2p-2}),
\end{eqnarray*}
where $\kappa_1=\{\odot_1(\mu_1,U_{11})\}^{2\tau}$, $\mu_1=M\bigr(c+\sum_{i=1}^{p}\phi_ix_{2i-1}+\sum_{j=1}^{p}\psi_jx_{2j}\bigr)$, $\phi_{i}=0$ for $p_1<i\leq p$, and $\psi_{j}=0$ for $p_2<j\leq p$.

The following proposition gives the conditions which ensure the
ergodicity and the stationarity of the RRC-GARCH$(p_1,p_2)$ process. For
$x\in\mathbb{E}$, and any measure $\lambda$
and function $g$ on $\mathbb{E}$, we set $\lambda(g)=\int
g(x)d\lambda(x)$.

\begin{proposition}\label{ergodicity}
Let $\psi(z)=1-\sum_{j=1}^{p_2}\psi_jz^i$ and $\varphi_*(z)=1-\sum_{j=1}^p(\phi_{j}^+ +\psi_{j}^+)z^{j}$, where $p=\max(p_1,p_2)$, $a^+=\max\{a,0\}$ for $a\in\mathbb{R}$, $\phi_{j}^+=0$ for $p_1<j\leq p$, and $\psi_{j}^+=0$ for $p_2<j\leq p$. Suppose that:
\begin{itemize}
\item[1.]The Markov chain $(\U_t)$ is irreducible and aperiodic;
\item[2.]For some $k \geq 2$, $E|\zeta_t|^k < +\infty$;
\item[3.]$\psi(z)\neq0$, $\varphi_*(z)\neq0$, for $z\in\mathbb{C}$ and $|z|\leq1$;
\item[4.]$0<\tau<1$.
\end{itemize}
Then
\begin{itemize}
\item[1.]The RRC-GARCH$(p_1,p_2)$ process $(\U_t)$ has a unique invariant probability measure $\lambda$;
\item[2.]For all $\u \in \mathbb{E}$ and $g \in L_1(\lambda)$, we have
$
\frac{1}{n}\sum_{t=1}^ng(\U_t)\longrightarrow \lambda(g),\ \
\mathbb{P}_{\text{\u}}\ a.s.
$
where $\mathbb{P}_{\text{\u}}$ denotes the conditional probability
$\mathbb{P}(\cdot)=\mathbb{P}(\cdot|\U_0=\u)$.
\item[3.] $E(X_t^k)<+\infty$.
\end{itemize}
\end{proposition}
Condition 1 is a necessary assumption for proving stationarity and
ergodicity of a Markov chain. Condition 2 is a standard moment condition to guarantee that the conditional mean and variance of $X_t$ exist. Condition 3 is an essential
condition for stationarity and ergodicity of a standard ARMA$(p,p_2)$
process. Note that, $\sum_{i=1}^{p_1}|\phi_i|+\sum_{j=1}^{p_2}|\psi_j|<1$ is a sufficient condition to ensure $\psi(z)\neq0$ and $\varphi_*(z)\neq0$ for $z\in\mathbb{C}$ and $|z|\leq1$. Condition 4 ensures the non-negative property of $X_{t}$, $t\in\mathbb{Z}$. Since the constraints on the parameters of a stationary ergodic
RRC-GARCH$(p_1,p_2)$ process are very weak, RRC-GARCH$(p_1,p_2)$ model can be
applied to analyze more real count data sets.

\begin{remark}\label{ergodicityremark}
The results in Proposition \ref{ergodicity} depend on $0<\tau<1$. From the proof of Proposition \ref{ergodicity}, we find that, the results in Proposition \ref{ergodicity} still hold when $\tau=1$, but require more stringent and complicated constraints on the parameters $\vtheta$ and $\vlambda$. We do not intend to pursue this direction further. Since it is difficult to verify these complicated constraints on the parameters in a real data analysis.
\end{remark}

\subsection{Autocorrelation structure}
Now, we show that the autocorrelation structure of a RRC-GARCH$(p_1,p_2)$ process with the proposed Laplace link function and $\vtheta\in\Theta_1$ is the same as that of a standard ARMA$(p,p_2)$.

Let $L_{\sigma+}(u)=L_\sigma(u){1}(u\geq0)=\{\sigma\log(2)+u\}{1}(u\geq0)$ and $L_{\sigma-}(u)=L_\sigma(u){1}(u<0)$. Then, define
\begin{eqnarray*}\label{muexpr}
\left\{
  \begin{array}{ll}
\xi_t=c+\sum_{i=1}^{p_1}\phi_iX_{t-i}+\sum_{j=1}^{p_2}\psi_j\mu_{t-j},\\
\mu_{t+}=L_{\sigma+}(\xi_t),\ \ \mu_{t-}=L_{\sigma-}(\xi_t),\\
\xi_t^+=L_0(\xi_t)\ \text{and}\ \xi_t^-=L_0(-\xi_t).
  \end{array}
\right.
\end{eqnarray*}
From (\ref{rinar0}) and (\ref{Laplace}) and $\xi_t=\xi_t^+-\xi_t^-$, it follows that
\begin{eqnarray*}
&&\mu_t=\mu_{t+}+\mu_{t-}=\sigma\log(2){1}(\xi_t\geq0)+\xi_t^+ +\mu_{t-}=\xi_t+\varrho_t,
\end{eqnarray*}
where
\begin{eqnarray*}
\varrho_t=\xi_t^-+\sigma\log(2){1}(\xi_t\geq0)+\mu_{t-}.
\end{eqnarray*}
Let $\phi(z)=\sum_{i=1}^{p_1}\phi_iz^i$ and recall that and $\psi(z)=1-\sum_{j=1}^{p_2}\psi_jz^i$. Then $\mu_t$ defined in (\ref{rinar0}) can be rewritten in terms of the backshift operator $B$ as $\psi(B)\mu_t=c+\phi(B)X_t+\varrho_t$. Note that $\psi(z)\neq0$ for $z\in\mathbb{C}$ and $|z|\leq1$ implies that $\psi^{-1}(z)=\text{inv}_{\psi}(z)=\sum_{i=0}^{+\infty}v_{\psi,i}z^i$ is well-defined for $|z|<1+\epsilon$ with some $\epsilon>0$, where $v_{\psi,i}$ is exponentially decreasing and defined recursively $v_{\psi,0}=1$ and $v_{\psi,n}=\sum_{i=1}^n\psi_iv_{\psi,n-i}$ for $n\geq1$. It follows that
\begin{eqnarray}\label{muexpr}
\mu_t&=&\frac{c}{\psi(1)}+\psi^{-1}(B)\phi(B)X_t+\psi^{-1}(B)\varrho_t=\theta_0+\theta(B)X_t+\psi^{-1}(B)\varrho_t,
\end{eqnarray}
where $\theta_0=\frac{c}{\psi(1)}$, $\theta(z)=\psi^{-1}(z)\phi(z)=\sum_{i=1}^{+\infty}\theta_iz_i$.

Based on (\ref{rinar1}), (\ref{eta}) and (\ref{muexpr}), we can write
\begin{eqnarray}\label{trans0}
X_t=\mu_t+\eta_t+\varepsilon_t=\frac{c}{\psi(1)}+\psi^{-1}(B)\phi(B)X_t+\psi^{-1}(B)\varrho_t+e_t,
\end{eqnarray}
where $e_t=\eta_t+\varepsilon_t$ and $\mu_t=\frac{c}{\psi(1)}+\psi^{-1}(B)\phi(B)X_t+\psi^{-1}(B)\varrho_t$. From (\ref{trans0}), we get
\begin{eqnarray*}\label{trans1}
\{\psi(B)-\phi(B)\}X_t=c+\varrho_t+\psi(B)e_t,
\end{eqnarray*}
which gives
\begin{eqnarray*}\label{trans2}
\biggr\{1-\sum_{j=1}^{p_2}\psi_jB^j-\sum_{i=1}^{p_1}\phi_iB^i\biggr\}X_t=c+\varrho_t+\biggr\{1-\sum_{j=1}^{p_2}\psi_jB^j\biggr\}e_t.
\end{eqnarray*}
Finally, we obtain
\begin{eqnarray}\label{trans3}
X_t&=&c+\varrho_t+\sum_{i=1}^{p_1}\phi_iX_{t-i}+\sum_{j=1}^{p_2}\psi_jX_{t-j}+e_t+\sum_{j=1}^{p_2}(-\psi_j)e_{t-j}\nonumber\\
&=&c+\varrho_t+\varphi(B)X_{t}+e_t+\delta(B)e_{t},
\end{eqnarray}
where $p=\max\{p_1,p_2\}$, $\varphi(B)=\sum_{k=1}^{p}\varphi_kB^k=\sum_{i=1}^{p_1}\phi_iB^i+\sum_{j=1}^{p_2}\psi_jB^j$, and $\delta(B)=\sum_{k=1}^{p_2}\delta_kB^k=\sum_{j=1}^{p_2}(-\psi_j)B^j$ with $\delta_k=-\psi_k$.

If $\vtheta\in\Theta_1$, then $\varrho_t\equiv \sigma\log(2)$ and the RRC-GARCH$(p_1,p_2)$ process can be written
\begin{eqnarray}\label{momentexp1}
X_t&=&\frac{c+\sigma\log(2)}{1-\varphi(1)}+\{1-\varphi(B)\}^{-1}\{1+\delta(B)\}e_t\nonumber\\
&=&\frac{c+\sigma\log(2)}{1-\varphi(1)}+\varpi(B)e_t=\frac{c+\sigma\log(2)}{1-\varphi(1)}+\sum_{i=0}^{+\infty}\varpi_ie_{t-i},
\end{eqnarray}
where $\varpi(B)=\{1-\varphi(B)\}^{-1}\{1+\delta(B)\}=\sum_{i=0}^{+\infty}\varpi_iB_i$ with $\varpi_0=1$.

Let $\mathscr{F} = (\mathcal{F}_n)_{n\in\mathbb{Z}}$ be the natural
filtration associated to the RRC-GARCH$(p_1,p_2)$ process, where
$\mathcal{F}_n =\sigma((U_{1t},U_{2t},\zeta_t), t \leq n)$
for $n\in\mathbb{Z}$, and $\mathcal{F}_{-\infty}$ is the degenerated
$\sigma$-algebra. It is easy to see that
$E(X_t|\mathcal{F}_{t-1})=\mu_t$ and
$\var(X_t|\mathcal{F}_{t-1})=R(\mu_t)+V_\tau(\mu_t) \sigma_\zeta^2 $.
From (\ref{trans3}) and (\ref{momentexp1}), we give the autocorrelation structure of the
RRC-GARCH$(p_1,p_2)$ process in the following proposition .

\begin{proposition}\label{autocorrelation}
Suppose that $\vtheta\in\Theta_1$ and the conditions of Proposition \ref{ergodicity} are
satisfied. Let
\begin{eqnarray*}
&&E(X_t)=\mu,\ \ \mbox{for all}\ t,\\
&&E(X_t-\mu)(X_{t-j}-\mu)=\gamma_j,\ \ \mbox{for all}\ t\ \mbox{and}\ j,\\
&&\rho_j=\gamma_j/\gamma_0,\ \ j\in\mathbb{Z}_+,
\end{eqnarray*}
then we have
\begin{eqnarray*}
\mu=\frac{c+\sigma\log(2)}{1-\sum_{i=1}^{p_1}\phi_i-\sum_{j=1}^{p_2}\psi_j}\ \ \text{and}\ \ \gamma_k=E(e_{t}^2)\sum_{i=0}^{+\infty}\varpi_i\varpi_{k+i},\ \ k\geq0,
\end{eqnarray*}
where $E(e_{t}^2)=E\{R(\mu_t)+\sigma_\zeta^2V_{\tau}(\mu_t)\}$ and $(\varpi_i)_{i\geq0}$ satisfies $\varpi(B)=\{1-\varphi(B)\}^{-1}\{1+\delta(B)\}=\sum_{i=0}^{+\infty}\varpi_iB_i$ with $\varpi_0=1$.
\end{proposition}

\subsection{Prediction}

For the RRC-GARCH type models, the one-step predictors of mean and
variance are given respectively by
\begin{eqnarray*}
E(X_{n+1}|\mathcal {F}_{n})=\mu_{n+1}\ \ \text{and}\ \ Var(X_{n+1}|\mathcal{F}_{n})=R(\mu_{n+1})+h_{n+1},
\end{eqnarray*}
where
\begin{eqnarray*}
h_{n+1}=D_\tau(\mu_{n+1})\sigma_\zeta^2,\\
\end{eqnarray*}
$\mu_{n+1}$ is defined in (\ref{rinar0}) and $D_\tau(\cdot)$ is given by (\ref{unifymvs}).

Note that the one-step predictors of mean and variance depend on
the unknown parameter vectors $\vtheta=(c,\phi_1,\cdots,\phi_{p_1},\psi_1,\cdots,\psi_{p_2})^\textsf{T}$ and $\vlambda=(\tau,\sigma_\zeta^2)^\textsf{T}$.  To use these these one-step predictors, we need to plug in consistent estimates of these two unknown parameter vectors. In the next
section, we will discuss the conditional least-squares estimation
for the unknown parameters.

\section{Conditional least-squares estimation}
\setcounter {equation}{0}
\def\theequation{\thesection.\arabic{equation}}
In this section, we state asymptotic results for the estimators of $\vtheta$
and $\vlambda$. We consider the conditional least-squares
estimator. The only requirement for an application of this method to
model (\ref{rinar1}) or the other RRC-GARCH type models is the ergodicity of the solution. But since
the Markov chain $(\U_t)$ is ergodic, the ergodicity of the process
$(X_t)$ easily follows.

Let $(W_t)$ be a stationary sequence of positive weights such that
$W_t\in\sigma(X_{t-1},\mu_{t-1},\cdots)$. For any $\vtheta\in\Theta_0$, define
\begin{eqnarray}\label{mutheta}
\mu_t(\vtheta)=M\bigr(\xi_t(\vtheta)\bigr)\ \text{and}\ \xi_t(\vtheta)=c+\sum_{i=1}^{p_1}\phi_iX_{t-i}+\sum_{j=1}^{p_2}\psi_j\mu_{t-j}(\vtheta),\ t\in
\mathbb{Z},
\end{eqnarray}
For model (\ref{rinar1}),
the weighted least-squares (WLS) estimator of $\vtheta$ is given by
\begin{eqnarray}\label{wls1}
\hat{\vtheta}_n=\arg\min_{\text{\b}\in\Theta}\frac{1}{n}\sum_{t=1}^nW_t\left\{X_t-\mu_{t}(\b)\right\}^2,
\end{eqnarray}
where $\vtheta\in\Theta$ and $\Theta$ is a compact set of $\Theta_0$. If $W_t\equiv1$, $\hat{\vtheta}_n$ is just the ordinary least squares (OLS)  estimator of $\vtheta$. In practice, minimization of (\ref{wls1}) can be done by an approximation procedure. Let $X_{t}=\tilde{\mu}_{t}(\vtheta)=0$ for $t\leq0$. Then, $\mu_t(\vtheta)$ can be approximated by
\begin{eqnarray*}\label{wls1approximate0}
\tilde{\mu}_t(\vtheta)=M\biggr(c+\sum_{i=1}^{p_1}\phi_iX_{t-i}+\sum_{j=1}^{p_2}\psi_j\tilde{\mu}_{t-j}(\vtheta)\biggr),\ t\in\mathbb{Z}.
\end{eqnarray*}
Correspondingly, $\hat{\vtheta}_n$ can be approximated by the solution of
$
\arg\min_{\text{\b}\in\Theta}\frac{1}{n}\sum_{t=1}^nW_t\left\{X_t-\tilde{\mu}_{t}(\b)\right\}^2.
$

\begin{theorem}\label{consistency}
Suppose that the conditions of Proposition \ref{ergodicity} are
satisfied. Furthermore, assume (i) $\mathbb{P}\{\mu_t(\vtheta)=\mu_t(\b)\}=1$ implies that $\vtheta=\b$; (ii)
\begin{eqnarray}\label{w1}
E\biggr(\sup_{\text{\b}\in\Theta}[W_t\{X_t-\mu_{t}(\b)\}^2]\biggr)<+\infty
\end{eqnarray}
Then, $\hat{\vtheta}_n$ is strongly consistent, i.e.
$\hat{\vtheta}_n\asrow\vtheta$.
\end{theorem}

\begin{theorem}\label{norm}
Suppose that the conditions of Theorem \ref{consistency} are
satisfied. Assume furthermore that the sequence of weights $(W_t)$
satisfies
\begin{eqnarray*}
&&E\biggr[\sup_{\text{\b}\in\mathcal {B}}\biggr\|W_t^{1/2}\frac{\partial \mu_{t}(\b)}{\partial \b}\biggr\|^2\biggr]<\infty,\ \ \lambda_{\min}\biggr[E\biggr\{W_t\frac{\partial \mu_{t}(\vtheta)}{\partial \vtheta}\frac{\partial \mu_{t}(\vtheta)}{\partial \vtheta^\textsf{T}}\biggr\}\biggr]>0,\\
&&E\biggr[\sup_{\text{\b}\in\Theta}\biggr\|W_t\{X_t-\mu_{t}(\b)\}\frac{\partial \mu_{t}(\b)}{\partial \b}\biggr\|^2\biggr]<\infty,\ \text{and}\ E\biggr\{W_t\biggr\|\frac{\partial \mu_{t}(\text{\vtheta})}{\partial \text{\vtheta}}\biggr\|^2\biggr\}<+\infty,
\end{eqnarray*}
where $\mathcal {B}\subseteq \Theta$ is a neighborhood of $\vtheta$ and $\lambda_{\min}(A)$ is the smallest eigenvalue of the matrix $A$. Then, the WLS estimator $\hat{\vtheta}_n$ is asymptotically normal,
i.e.
\begin{eqnarray*}
\sqrt{n}(\hat{\vtheta}_n-\vtheta)\longrightarrow
N(0,K_1^{-1}\Gamma_1 K_1^{-1}),
\end{eqnarray*}
where
\begin{eqnarray*}
K_1=E\biggr\{W_t\frac{\partial \mu_{t}(\vtheta)}{\partial \vtheta}\frac{\partial \mu_{t}(\vtheta)}{\partial \vtheta^\textsf{T}}\biggr\},\ \ \ \ \
\Gamma_1=E\biggr[W_t^2\{X_t-\mu_{t}(\vtheta)\}^2\frac{\partial \mu_{t}(\vtheta)}{\partial \vtheta}\frac{\partial \mu_{t}(\vtheta)}{\partial \vtheta^\textsf{T}}\biggr].
\end{eqnarray*}
\end{theorem}

\begin{remark}
The integrability conditions on the weights ensure that the matrices
$K_1$ and and $\Gamma_1$ are well defined. It is well-known that the
optimal choice of the weights for the asymptotic variance in the WLS
estimation is given by
\begin{eqnarray}\label{conditionvar}
W_t=Var(X_t|\mathcal
{F}_{t-1})^{-1}=\{R(\mu_t)+h_t\}^{-1},\
\ t\in\mathbb{Z}_+
\end{eqnarray}
where
\begin{eqnarray*}
h_{t}=h_{t}(\vlambda|\vtheta)=\sigma_\zeta^2D_\tau(\mu_{t}(\vtheta)),
\end{eqnarray*}
and $\mu_t=\mu_t(\vtheta)$ is defined by (\ref{mutheta}).
\end{remark}
If we want to apply the optimal weights in (\ref{conditionvar}) to obtain a more efficient estimator of $\vtheta$, we
have to replace the unknown parameter vectors $(\vtheta^\textsf{T},\vlambda^\textsf{T})$ with an initial estimator, e.g., the OLS estimator
$(\hat{\vtheta}^\textsf{T},\hat{\vlambda}^\textsf{T})$. The following theorem
justifies this two-steps procedure.

\begin{theorem}\label{twostep}
Let
$(\hat{\vtheta}^\textsf{T},\hat{\vlambda}^\textsf{T})$ be a sequence of estimators such that
$\sqrt{n}((\hat{\vtheta}-\vtheta)^\textsf{T},(\hat{\vlambda}-\vlambda)^\textsf{T})=O_{\mathbb{P}}(1)$.
Suppose that the conditions of Proposition \ref{ergodicity} are satisfied. Moreover, assume
\begin{eqnarray*}
E\biggr[\sup_{\|\text{\vtheta}-\text{\b}\|+\|\text{\vlambda}-\text{\s}\|\leq \delta_n}\sup_{1\leq t \leq n}|h_{t}(\vlambda|\vtheta)-h_{t}(\s|\b)|\biggr]<+\infty,
\end{eqnarray*}
where $\delta_n>0$ and $\lim_{n\rightarrow+\infty}\delta_n=0$. Define the optimal WLS (OWLS) estimator of $\vtheta$ as
\begin{eqnarray}\label{wlsoptimal}
\breve{\vtheta}_n=\arg\min_{\text{\b}\in\Theta}\frac{1}{n}\sum_{t=1}^n\hat{W}_t\left\{X_t-\mu_{t}(\b)\right\}^2.
\end{eqnarray}
where
$\hat{W}_t=\left(R(\hat{\mu}_t)+\hat{h}_t\right)^{-1}$ with
$\hat{h}_{t}=h_{t}(\hat{\vlambda}|\hat{\vtheta})=\hat{\sigma}_\zeta^2D_{\hat{\tau}} (\mu_t(\hat{\vtheta}))$ and $\hat{\mu}_t=\mu_t(\hat{\vtheta})$.
Then, we have
$
\sqrt{n}(\breve{\vtheta}_n-\vtheta)\longrightarrow N(0,\Sigma),
$
where
$$
\Sigma^{-1}=E\left\{\left(R(\mu_t)+h_t\right)^{-1}\frac{\partial \mu_{t}(\vtheta)}{\partial \vtheta}\frac{\partial \mu_{t}(\vtheta)}{\partial \vtheta^\textsf{T}}\right\}.
$$
\end{theorem}
For RRC-GARCH type models, OLS estimator of $\vlambda$ is given
by
\begin{eqnarray}\label{wlsv}
\hat{\vlambda}_n=(\hat{\tau}_n,\hat{\sigma}_{n\zeta}^2)=\arg\min_{\text{\s}\in\Lambda}\frac{1}{n}\sum_{t=1}^n\left[\{X_t-\mu_t(\hat{\vtheta}_n)\}^2-R(\mu_t(\hat{\vtheta}_n))-h_t(\s|\hat{\vtheta}_n)\right]^2,
\end{eqnarray}
where $\vlambda\in\Lambda$ and $\Lambda$ is a compact set of $(0,1]\times \mathbb{R}_+$ and  $\hat{\vtheta}_n$ is the OLS estimator of $\vtheta$.

To solve $\hat{\vlambda}_n$, for fixed $\tau$, define $$\sigma_{n\zeta}^2(\tau)=\arg\min_{\sigma_{\zeta}^2>0}\frac{1}{n}\sum_{t=1}^n\left[\{X_t-\mu_t(\hat{\vtheta}_n)\}^2-R(\mu_t(\hat{\vtheta}_n))-h_t(\s|\hat{\vtheta}_n)\right]^2.$$ It is easy to see that
\begin{eqnarray*}\label{ols2approximatevartheta}
\sigma_{n\zeta}^2(\tau)=\left(\frac{1}{n}\sum_{t=1}^nD_\tau^2(\mu_{t}(\hat{\vtheta}_n))\right)^{-1}\frac{1}{n}\sum_{t=1}^nD_\tau(\mu_{t}(\hat{\vtheta}_n)) [\{X_t-\mu_t(\hat{\vtheta}_n)\}^2-R(\mu_t(\hat{\vtheta}_n))].
\end{eqnarray*}
The estimator $(\hat{\tau}_n,\hat{\sigma}_{n\zeta}^2)$ can be computed by
\begin{eqnarray*}\label{ols2approximategamma}
\hat{\tau}_n&=&\arg\min_{0<\tau\leq1}\frac{1}{n}\sum_{t=1}^n\left[\{X_t-\mu_t(\hat{\vtheta}_n)\}^2-R(\mu_t(\hat{\vtheta}_n))-\sigma_{n\zeta}^2(\tau)D_\tau(\mu_{t}(\hat{\vtheta}_n))\right]^2,\\
\hat{\sigma}_{n\zeta}^2&=&\sigma_{n\zeta}^2(\hat{\tau}_n).
\end{eqnarray*}

We have the following result:
\begin{theorem}\label {vartheconsist}
Suppose that the conditions of Theorem \ref{norm} are satisfied. Furthermore, assume that (i) $\mathbb{P}\{h_t(\vlambda|\vtheta)=h_t(\s|\vtheta)\}=1$ implies that $\vlambda=\s$; (ii) $E(\sup_{\text{\b}\in\mathcal {B},\s\in\Lambda}[\{X_t-\mu_t(\b)\}^2-R(\mu_t(\b))-h_t(\s|\b)]^2)<+\infty$, where $\mathcal {B}\subseteq \Theta$ is a neighborhood of $\vtheta$. Then,
$\hat{\vlambda}_n$ is weakly consistent, i.e. $\hat{\vlambda}_n\prow \vlambda$.
\end{theorem}

\begin{theorem}\label {varthenormal}
Suppose that the conditions of Theorem \ref{vartheconsist} are
satisfied. Let $u_t(\vlambda|\vtheta)=\{X_t-\mu_t(\vtheta)\}^2-R(\mu_t(\vtheta))-h_t(\vlambda|\vtheta)$. Furthermore, assume that
\begin{eqnarray*}
&&E\biggr\{\sup_{\text{\b}\in\mathcal {B},\s\in\Lambda}\biggr\|\frac{\partial h_t(\text{\s}|\text{\b})}{\partial\text{\s}}\biggr\|^2\biggr\}<+\infty,\ \  E\biggr\{\sup_{\text{\b}\in\mathcal {B},\s\in\Lambda}\biggr\|\frac{\partial h_t(\text{\s}|\text{\b})}{\partial\text{\s}}\biggr\|\biggr\|\frac{\partial u_t(\text{\s}|\text{\b})}{\partial \text{\b}^\textsf{T}}\biggr\|\biggr\}<+\infty,\\
&&\lambda_{\min}\biggr[E\biggr\{\frac{\partial h_t(\vlambda|\vtheta)}{\partial\vlambda}\frac{\partial u_t(\vlambda|\vtheta)}{\partial \vlambda^\textsf{T}}\biggr\}\biggr]>0,\ \text{and}\ E\biggr\{\biggr\|\frac{\partial h_t(\vlambda|\vtheta)}{\partial\vlambda}u_t(\vlambda|\vtheta)\biggr\|^2\biggr\}<+\infty.
\end{eqnarray*}
Then, the OLS
estimator $\hat{\vlambda}_n$ is asymptotically normal, i.e.
\begin{eqnarray*}
\sqrt{n}(\hat{\vlambda}_n-\vlambda)\longrightarrow
N(0,\Omega),
\end{eqnarray*}
where $\Omega$ is given in the proof of this theorem.
\end{theorem}
The algorithms for the computations of the OLS and OWLS estimates and their estimated asymptotic covariance matrices
are provided in the Section S2 of the Supplementary Material.

\section{Model selection and model diagnostics}
\setcounter {equation}{0}
\def\theequation{\thesection.\arabic{equation}}

\subsection{Model selection}
In this subsection, we consider the model selection problem for the
RRC-GARCH models. Recall that $\hat{\vtheta}_n$ is the OLS estimator of $\vtheta$ and $\hat{\vlambda}_n=(\hat{\tau}_n,\hat{\sigma}_{n\zeta}^2)$
is the OLS estimator of $\vlambda$. Without specifying the
distribution of $\zeta_t$ in (\ref{rinar1}), the conditional
likelihood function of the RRC-GARCH model can not be available. Following the idea
of Hurvich \& Tsai (1995) and Liu \& Yuan (2013), we propose to use the conditional
Gaussian quasi-likelihood to construct AIC and BIC for RRC-GARCH$(p_1,p_2)$
models. The resultant AIC and BIC are given respectively as
\begin{eqnarray*}
\mbox{AIC}(p_1,p_2)&=&\sum_{t=1}^n\log\left\{R(\hat{\mu}_t)+\hat{\sigma}_{n\zeta}^2D_{\hat{\tau}_n} (\hat{\mu}_t)\right\}+2(3+p_1+p_2),
\end{eqnarray*}
and
\begin{eqnarray*}
\mbox{BIC}(p_1,p_2)&=&\sum_{t=1}^n\log\left\{R(\hat{\mu}_t)+\hat{\sigma}_{n\zeta}^2D_{\hat{\tau}_n} (\hat{\mu}_t)\right\}+\log(n-p-1)(3+p_1+p_2),
\end{eqnarray*}
where $\hat{\mu}_t=\mu_t(\hat{\vtheta}_n)$ and $\mu_t(\vtheta)$ is given in (\ref{mutheta}).

Let $(p_{1m},p_{2m})$ be a maximum model order cut-offs for RRC-GARCH$(p_1,p_2)$
model. The selected order using AIC and BIC is given respectively by
\begin{eqnarray*}
(\hat{p}_1,\hat{p}_2)_{AIC}=\arg\min_{p_k\leq p_{km},k=1,2}\mbox{AIC}(p_1,p_2),
\end{eqnarray*}
and
\begin{eqnarray*}
(\hat{p}_1,\hat{p}_2)_{BIC}=\arg\min_{p_k\leq p_{km},k=1,2}\mbox{BIC}(p_1,p_2).
\end{eqnarray*}
We investigate the performances of these two criteria for selecting the RRC-GARCH
models through simulations in the next section.

\subsection{Model diagnostics}

To check the adequacy of the conditional moments assumptions of RRC-GARCH models, we consider the standardized Pearson residuals $(r_t)$, which are given by
\begin{eqnarray*}
r_t(\vtheta_n,\vlambda_n)=\frac{X_t-E(X_t|\mathcal {F}_{t-1};\vtheta_n,\vlambda_n)}{\sqrt{\var(X_t|\mathcal {F}_{t-1};\vtheta_n,\vlambda_n)}}=\frac{X_t-\mu_t(\vtheta_n)}{\sqrt{R(\mu_t(\vtheta_n))+V_{\tau_n}(\mu_t(\vtheta_n)) \sigma_{n\zeta}^2 }},
\end{eqnarray*}
where $\vlambda_n=(\tau_n,\sigma_{n\zeta}^2)^\textsf{T}$ is the OLS estimate of $\vlambda$ and $\vtheta_n$ is the OLS or OWLS estimate of $\vtheta$. For an adequate RRC-GARCH model, the Pearson residuals should have mean zero, variance one, and be uncorrelated. In particular, Aknouche \& Scotto (2024) used the mean absolute residual (MAR), $\text{MAR}=\frac{1}{n}\sum_{t=1}^n|X_t-\mu_t(\vtheta_n)|$, and the mean squared Pearson residual, $\text{MSPR}=\frac{1}{n}\sum_{t=1}^nr_t^2(\vtheta_n,\vlambda_n)$, for checking the adequacy of the dispersion structure. Obviously, smaller values of MAR and $|\text{MSPR}-1|$ indicate a better model fit.

\section{Simulation experiment}
\setcounter {equation}{0}
\def\theequation{\thesection.\arabic{equation}}
To evaluate the efficiency of the conditional least-squares estimators and the performances of the proposed AIC and BIC for
selecting the RRC-GARCH models with the Laplace link function $L_1(\cdot)$, we conduct two simulation studies.
All simulations are carried out in the R Project for Statistical Computing. Moreover, in all
simulations, the innovation variable, say $\zeta_t$, is generated from a Binomial
distribution with parameter $(2,1/2)$. Moreover, we set $\tau=0.5$.  Additional simulation results with $\tau=0.2$ or 0.8 presented in the Section S3 of the Supplementary Material yield similar
results.

For the two Monte Carlo studies, we consider the following two settings:
\begin{itemize}
\item[]Setting (a):
\item[$M_1$:] RRC-GARCH(1,0) model with   $(\vtheta^\textsf{T},\vlambda^\textsf{T} )=(-0.4,0.5,0.5,0.5)$;
\item[$M_2$:] RRC-GARCH(1,1) model with  $(\vtheta^\textsf{T},\vlambda^\textsf{T} )=(-0.4,0.4,0.4,0.5,0.5)$;
\item[$M_3$:] RRC-GARCH(1,2) model with  $(\vtheta^\textsf{T},\vlambda^\textsf{T} )=(-0.4,0.4,0.1,0.4,0.5,0.5)$;
\item[$M_4$:] RRC-GARCH(2,0) model with    $(\vtheta^\textsf{T},\vlambda^\textsf{T} )=(-0.4,0.2,0.5,0.5,0.5)$;
\item[$M_5$:] RRC-GARCH(2,1) model with   $(\vtheta^\textsf{T},\vlambda^\textsf{T} )=(-0.4,0.1,0.4,0.4,0.5,0.5)$;
\item[$M_6$:] RRC-GARCH(2,2) model with   $(\vtheta^\textsf{T},\vlambda^\textsf{T} )=(-0.4,0.1,0.4,0.1,0.3,0.5,0.5)$;
\end{itemize}
\begin{itemize}
\item[]and Setting (b):
\item[$M_1$:] RRC-GARCH(1,0) model with   $(\vtheta^\textsf{T},\vlambda^\textsf{T} )=(2,-0.5,0.5,0.5)$;
\item[$M_2$:] RRC-GARCH(1,1) model with  $(\vtheta^\textsf{T},\vlambda^\textsf{T} )=(2,-0.4,-0.4,0.5,0.5)$;
\item[$M_3$:] RRC-GARCH(1,2) model with  $(\vtheta^\textsf{T},\vlambda^\textsf{T} )=(2,-0.4,-0.1,-0.4,0.5,0.5)$;
\item[$M_4$:] RRC-GARCH(2,0) model with    $(\vtheta^\textsf{T},\vlambda^\textsf{T} )=(2,-0.2,-0.5,0.5,0.5)$;
\item[$M_5$:] RRC-GARCH(2,1) model with   $(\vtheta^\textsf{T},\vlambda^\textsf{T} )=(2,-0.1,-0.4,-0.4,0.5,0.5)$;
\item[$M_6$:] RRC-GARCH(2,2) model with   $(\vtheta^\textsf{T},\vlambda^\textsf{T} )=(2,-0.1,-0.4,-0.1,-0.3,0.5,0.5)$.
\end{itemize}

To get an intuition about the abilities of the RRC-GARCH models with the Laplace link function $L_1(\cdot)$ for explaining different autocorrelation structures, we present sample ACF pairs
$(\rho_{X_t}(1),\rho_{X_t}(2))$ for the above six RRC-GARCH models in Table 1. From Table 1, we see that, the RRC-GARCH models can exhibit different patterns of autocorrelation structures.

In the first simulation study, we use the root mean squared error (RMSE) to evaluate the finite sample behaviour of the conditional least squares estimators.  We consider two sample sizes: $n=200,\ 500$ and the number of replications is set to be $1000$.  Simulation results for the settings (a) and (b) are
presented in Tables 2-3, respectively. We find that the OWLS estimator $\breve{\vtheta}_n$
gives smaller RMSEs for $\vtheta$ than the OLS estimator $\hat{\vtheta}_n$ in most
cases. Thus, the finite sample performance and the large sample asymptotic theory both show that the OWLS estimator is more
efficient than the OLS estimator.

In the second Monte Carlo study, we examine the performances of AIC
and BIC for selecting the RRC-GARCH models with the Laplace link function $L_1(\cdot)$. We set the maximum model order cut-offs
$(p_{1m}, p_{2m})$ as (2,2). Thus, the set of candidate models is $\{M_j: j=1,\cdots,6\}$.

We consider three sample sizes: $n=100$, $200$ and $500$. Tables 4-5 give the numbers of the order selected by AIC
and BIC in 1000 realizations. From the results reported in Tables 4-5, we find that AIC performs better than BIC in
most cases for $n=100$ and $200$, but BIC outperforms AIC in most cases
when $n=500$. Obviously, BIC tends to select a simpler model with small sample sizes while AIC tends to select a larger model. As the sample size $n$ increases from 100 to 500, both these two criteria perform very well according to the high probability for selecting the true model.

\section{Applications to real data}
\setcounter {equation}{0}
\def\theequation{\thesection.\arabic{equation}}
To demonstrate how the RRC-GARCH models work, they are applied to two data sets in completely different areas.

\subsection{Disease counts}
In this subsection, we consider the weekly counts of disease cases, which are caused by Escherichia coli (Ecoli) and reported for North Rhine-Westphalia (Germany) from January 2001 to May 2013.
These data were originally taken from SurvStat@RKI 2.0 at \textit{https://survstat .rki .de/} by Liboschik et al. (2017) and can be found via the command \verb"ecoli" of the R-package \textsc{tscount}.

The length of the series is $646$. The counts vary from 3 to 92 and its mean and variance is 20.3344 and 88.7531, respectively. Figure 4 (a, b, c) shows the plots of this time series, its sample autocorrelation (ACF) and sample partial autocorrelation (PACF). From Figure 4 (b,c), the PACF graph is truncated at the 2-th order, while the ACF graph is trailing. The sample ACF and PACF imply that the RRC-GARCH(2,0) model should be considered.

We use the first $n=616$ observations for fitting the RRC-GARCH models with the Laplace link function $L_1(\cdot)$ and leave out the last $n_{new}=30$ observations for
a later forecast experiment. For model selection, we consider RRC-GARCH$(p_1,p_2)$ models with $(p_1,p_2)=(1,0)$, $(1,1)$, $(1,2)$, $(2,0)$, $(2,1)$, and $(2,2)$ as candidate models. The AIC and BIC values of the candidate models for the Ecoli counts are given in Table 6. It is easy to see that both of these two criteria select the
RRC-GARCH(2,0) model.

The final estimates together with their estimated standard deviations (SDs) are summarized in Table 7. Obviously, the OWLS estimate $\breve{\vtheta}_n\in \Theta_1$
and the OLS estimate $\hat{\vtheta}_n\in \Theta_1$, which implies that the fitted RRC-GARCH(2,0) model has a linear mean structure. Moreover, the OWLS estimator $\breve{\vtheta}_n$ gives smaller SDs for $\vtheta$ than the OLS estimator $\hat{\vtheta}_n$.

Figure 4(d) presented the standardized residuals plot using the OLS estimate of $\vtheta$. To check the adequacy of the fitted models, we consider the approaches discussed in Section 5.2. The model diagnostics statistics MAR and MSPR with the OLS and OWLS estimates of $\vtheta$ are presented in Table 8. The sample means ($E_n(r_t)$), sample SDs ($Sd_n(r_t)$), and the maximum absolute value of the sample autocorrelation ($\max_{1\leq k\leq 20}|\rho_{r_t}(k)|$) of the standardized residuals using the OLS and OWLS estimates of $\vtheta$ are also reported. From Table 8, we see that, the OLS and OWLS methods have the same MAR values but the MSPR value of the OWLS method is closer to 1 than that of the OLS method using the RRC-GARCH(2,0) model, which indicates a better model fit.

Finally, let us analyze the forecast performance of the fitted RRC-GARCH(2,0) model with the OLS and OWLS estimates of $\vtheta$. We apply the MAR and MSPR criteria to the 30 new Ecoli counts. The sample means ($E_{n_{new}}(r_t)$), sample SDs ($Sd_{n_{new}}(r_t)$), and the maximum absolute value of the sample autocorrelation ($\max_{1\leq k\leq 14}|\rho_{r_t}(k)|$) of the standardized prediction residuals using the OLS and OWLS estimates of $\vtheta$ are also considered. Results are summarized in Table 9.  Obviously, the OWLS-fitted RRC-GARCH(2,0) model shows the better predictive performance regarding the 30 Ecoli counts, according to the MAR and MSPR criteria.

\subsection{Transaction counts}

In the following, we shall investigate financial transactions counts data. Aknouche et al. (2022) provided one of the time series, i.e., the number of stock transactions concerning
the Wausau Paper Corporation (WPP), measured in 5-min intervals between 9:45 AM and 4:00 PM for the period from January 3 to
February 18 in 2005.

The length of the series is $2925$. The counts vary from 0 to 43 and its mean and variance is 8.1115 and 35.5382, respectively. Figure 5 (a, b, c) shows the plots of this time series, its sample autocorrelation (ACF) and sample partial autocorrelation (PACF). From Figure 5 (b,c), the PACF and ACF graphs are trailing. The sample ACF and PACF imply that the RRC-GARCH$(p_1,p_2)$ models with $p_1\geq1$ and $p_2\geq1$ should be considered.

In analogy to Aknouche et al. (2022, Section 6.2) and  Wei${\ss}$ \& Zhu (2024, Section 6.2), we use the first $n=2825$ observations for fitting the RRC-GARCH models with the Laplace link function $L_1(\cdot)$ and leave out the last $n_{new}=100$ observations for a later forecast experiment. For model selection, we consider RRC-GARCH$(p_1,p_2)$ models with $(p_1,p_2)=(1,0)$, $(1,1)$, $(1,2)$, $(2,0)$, $(2,1)$, and $(2,2)$ as candidate models. The AIC and BIC values of the candidate models for the WPP counts are given in Table 6. Obviously, both of these two criteria select the RRC-GARCH(2,1) model.

The final estimates together with their estimated standard deviations (SDs) are summarized in Table 7. Obviously, the OWLS estimator $\breve{\vtheta}_n\notin \Theta_1$
and the OLS estimator $\hat{\vtheta}_n\notin \Theta_1$, which implies that the fitted RRC-GARCH(2,1) model has a nonlinear mean structure. Moreover, the OWLS estimator $\breve{\vtheta}_n$ gives smaller SDs for $\vtheta$ than the OLS estimator $\hat{\vtheta}_n$.

Figure 5(d) presents the standardized residuals plot using the OLS estimates of $\vtheta$ of $\vlambda$. To check the adequacy of the fitted models, we also consider the approaches discussed in Section 5.2. The model diagnostics statistics MAR and MSPR with the OLS and OWLS estimates of $\vtheta$ are presented in Table 8. The sample means ($E_n(r_t)$), sample SDs ($Sd_n(r_t)$), and the maximum absolute value of the sample autocorrelation ($\max_{1\leq k\leq 20}|\rho_{r_t}(k)|$) of the standardized residuals using the OLS and OWLS estimates of $\vtheta$ are also reported. From Table 8, we see that, the OLS and OWLS methods have the same MAR values and similar MSPR values using the RRC-GARCH(2,1) model, which indicates a similar model fit.

Finally, let us evaluate the forecast performance of the fitted RRC-GARCH(2,1) model with the OLS and OWLS estimates of $\vtheta$. We apply the MAR and MSPR criteria to the 100 new WPP counts. The sample means ($E_{n_{new}}(r_t)$), sample SDs ($Sd_{n_{new}}(r_t)$), and the maximum absolute value of the sample autocorrelation ($\max_{1\leq k\leq 14}|\rho_{r_t}(k)|$) of the standardized prediction residuals using the OLS and OWLS estimates of $\vtheta$ are also considered. Results are summarized in Table 9.  Obviously, the OWLS-fitted RRC-GARCH(2,1) model shows the better predictive performance than the OLS-fitted RRC-GARCH(2,1) model regarding the 100 WPP counts, in terms of the MAR and MSPR criteria. Compared to the fitted CMEM model (MAR=3.613, MSPR=1.164) with Poi-counting series in Wei${\ss}$ \& Zhu (2024), the OWLS-fitted RRC-GARCH(2,1) model (MAR=3.5785, MSPR=1.1848) gives a smaller MAR value but a larger MSPR value. In summary, the OWLS-fitted RRC-GARCH(2,1) model shows the best predictive performance regarding the last 100 WPP counts, in terms of the MAR criterion.

\section{Discussion}
\setcounter {equation}{0}
\def\theequation{\thesection.\arabic{equation}}
In this paper, we developed RRC-GARCH models for the analysis of count-valued time series. The RRC-GARCH model and its variants can provide flexible and feasible MVSs.  The new model using the proposed Laplace link functions with an appropriate parameter space has a flexible range of ACF values and exhibits a linear mean structure, which makes its model parameters easier to interpret than those of a pure non-linear mean model.  The OLS and OWLS estimators were used to estimate the model parameters, and their large-sample properties were derived. The proposed RRC-GARCH model offers a promising approach to jointly model the conditional mean and variance of count data. Its flexibility in handling different mean-variance structures and ability to provide efficient forecasts make it a valuable tool for analyzing count data.

In this paper, we only consider the conditional least-squares estimators of the regression parameters $(\vtheta,\vlambda)$ in the conditional mean and variance of the RRC-GARCH models, and the distribution of $\zeta_t$ is not specified and remains nonparametric. We may consider the non-parametric maximum likelihood estimators (NPMLE) of $(\vtheta,\vlambda)$ and the distribution of $\zeta_t$ in the RRC-GARCH model. Then, the NPMLE-based estimator of the forecast distribution may be obtained.

\section*{Supplementary Material}

The online Supplementary Material contains the proofs of all Theorems, the detailed computation algorithm for the OLS and OWLS estimates and their estimated asymptotic covariance matrices, and additional simulation results.

\section*{Acknowledgements}

We thank Professors Danning Li and Lianyan Fu for help discussions.


\makeatletter
\renewenvironment{thebibliography}[1]
{\section*{\refname}%
\@mkboth{\MakeUppercase\refname}{\MakeUppercase\refname}%
\list{\@biblabel{\@arabic\c@enumiv}}%
{\settowidth\labelwidth{\@biblabel{#1}}%
\leftmargin\labelwidth \advance\leftmargin\labelsep
\advance\leftmargin by 2em%
\itemindent -2em%
\@openbib@code
\usecounter{enumiv}%
\let\p@enumiv\@empty
\renewcommand\theenumiv{\@arabic\c@enumiv}}%
\sloppy \clubpenalty4000 \@clubpenalty \clubpenalty
\widowpenalty4000%
\sfcode`\.\@m} {\def\@noitemerr
{\@latex@warning{Empty `thebibliography' environment}}%
\endlist}
\renewcommand\@biblabel[1]{}
\makeatother

\begin{thebibliography}{}


\bibitem{Aknouche2022}
Aknouche, A., Almohaimeed, B. S. \& Dimitrakopoulos, S. (2022) Forecasting transaction counts with integer-valued GARCH models. {\it Studies in Nonlinear Dynamics \& Econometrics} \textbf{26}:
529--539.

\bibitem{Aknouche2024}
Aknouche, A. \& Scotto, M. G. (2024) A multiplicative thinning-based integer-valued GARCH model. {\it  Journal of Time Series Analysis}  \textbf{45}: 4--26.


\bibitem{Billingsley1968}
 Billingsley, P. (1968) {\it Convergence of Probability Measures}. New
York: John Wiley.

\bibitem{Box1976}
Box, G. E. P. (1976) Science and statistics. {\it Journal of the American Statistical Association} \textbf{71 }(356): 791--799.

\bibitem{Cameron2013}
Cameron, A. C. \& Trivedi, P. K. (2013) {\it Regression Analysis of Count Data.}  Cambridge, UK: Cambridge Univ.
Press.

\bibitem{Chen2022}
Chen, H., Li, Q. \& Zhu, F. (2022) A new class of integer-valued GARCH models for time series of bounded
counts with extra-binomial variation. {\it AStA Advances in Statistical Analysis} \textbf{106}: 243--270.


\bibitem{Davis2000}
Davis, R.~A., Dunsmuir, W.~T.~M. \&  Wang, Y.  (2000) On
autocorrelation in a Poisson regression model. {\it Biometrika}
\textbf{87}, 491--505.


\bibitem{Davis2003}
Davis, R.~A., Dunsmuir,  W.~T.~M. \&  Streett, S.~B. (2003)  Observation-driven
models for Poisson counts. {\it Biometrika} \textbf{90}, 777--790.

\bibitem{Davis2009}
Davis, R.~A. \& Wu, R.  (2009) A negative binomial model for time
series of counts. {\it Biometrika} \textbf{96}, 735--749.

\bibitem{Dugas2000}
Dugas, C., Bengio, Y., B\'{e}lisle, F., Nadeau, C. \& Garcia, R. (2000) Incorporating second-order
functional knowledge for better option pricing. In Proceedings of the 13th International Conference on Neural Information Processing Systems (NIPS'00) (Edited by Leen et al.),
451--457. MIT Press, Cambridge.


\bibitem{Heinen2003}
 Heinen, A. (2003) {\it Modeling Time Series Count Data: An Autoregressive Conditional
Poisson Model.} CORE Discussion Paper 2003/62, Universit\'{e}
catholique de Louvain.


\bibitem{Hurvich1995}
 Hurvich, C.~M. \&   Tsai, C. (1995) Model selection for extended quasi-likelihood models in small
samples. {\it Biometrics} \textbf{51},   1077--1084.

\bibitem{Jia2023}
Jia, Y., Kechagias, S., Livsey, J., Lund, R. \& Pipiras, V. (2023) Latent Gaussian count time series.
{\it Journal of the American Statistical Association} \textbf{118}(541): 596--606.


\bibitem{Karlis2023}
Karlis, D. \& Mamode Khan, N. M. (2023) Models for Integer Data. {\it Annual Review of Statistics and Its Application}
\textbf{10}: 297--323.


\bibitem{Kedem2002}
 Kedem, B. \&   Fokianos, K. (2002) {\it Regression Models for Time Series
Analysis.}  Hoboken, New Jersey: Wiley.

\bibitem{Gorgi2020}
Gorgi, P. (2020) Beta-negative binomial auto-regressions for modelling integer-valued time series with extreme
observations. {\it Journal of the Royal Statistical Society, Series B (Statistical Methodology)} \textbf{82}(5): 1325--1347.


\bibitem{Kong2003}
Kong, J. \& Lund, R. (2023) Seasonal count time series. {\it Journal of Time Series Analysis} \textbf{44}(1): 93--124.


\bibitem{Li2024}
Li, Q., Chen, H. \& Zhu, F. (2024) $\mathbb{Z}$-valued time series: Models, properties and comparison. {\it Journal of Statistical Planning and Inference} \textbf{230}:106099

\bibitem{Liboschik2017}
Liboschik, T., Fokianos, K. \& Fried, R. (2017) tscount: an R package for analysis of count time series following generalized linear models. {\it Journal of Statistical Software } \textbf{82}: 1--51.


\bibitem{Liu2013}
Liu, T. \& Yuan, X. (2013) Random rounded integer-valued autoregressive conditional heteroskedastic process. {\it Statistical Papers} \textbf{54}: 645--683.

\bibitem{Maya2022}
Maya, R., Chesneau, C., Krishna, A. \& Irshad, M. R. (2022) Poisson
extended exponential distribution
with associated INAR(1) process and
applications. {\it Statistics } \textbf{5}: 755-772.


\bibitem{McKenzie1985}
McKenzie, E. (1985). Some simple models for discrete variate time series. {\it Water Resources Bulletin} \textbf{21}(4), 645--650.


\bibitem{McKenzie2003}
 McKenzie, E. (2003) Discrete variate time series. In {\it Handbook of
 Statistics 21: Stochastic Processes:
Modeling and Simulation} (eds C. R. Rao and D. N. Shanbhag).
Amsterdam: Elsevier Science, pp. 573--606.

\bibitem{Mei2017}
Mei, H. \& Eisner, J. (2017) The neural Hawkes process: A neurally self-modulating multi-
variate point process. In Proceedings of the 31st International Conference on Neural Infor-
mation Processing Systems (NIPS'17) (Edited by von Luxburg et al.), Curran Associates
Inc., 6757--6767.


\bibitem{Meyn2009}
Meyn, S. \& Tweedie, R. L. (2009) Markov Chains and Stochastic Stability, second ed.
Cambridge University Press, Cambridge.

\bibitem{Newey1994}
Newey, W. K. \& McFadden, D. (1994) Large sample estimation and hypothesis testing, in: R. Engle, D. McFadden (Eds.), Handbook of Econometrics, Vol. IV,
Elsevier, Amsterdam, pp. 2111-2245.



\bibitem{Scotto5015}
Scotto, M. G., Wei${\ss}$, C.~H. \& Gouveia, S. (2015) Thinning-based models in the analysis of
integer-valued time series: A review.  {\it Statistical Modelling} \textbf{15}(6): 590--618.


\bibitem{Wei2018}
 Wei${\ss}$, C.~H. (2018) {\it An Introduction to Discrete-Valued Time Series.} John Wiley \& Sons, Chichester.

\bibitem{Wei2024}
 Wei${\ss}$, C.~H. \& Zhu, F.  (2024) Conditional-mean multiplicative operator models for count time
series.  {\it Computational Statistics and Data Analysis}  \textbf{191}, 107885

\bibitem{Wei2022}
 Wei${\ss}$, C.~H., Zhu, F. \& Hoshiyar, A. (2022) Softplus INGARCH models  {\it Statistica Sinica}  \textbf{32}, 1099--1120.

\bibitem{Zheng2015}
Zheng, T., Xiao, H. \& Chen, R. (2015) Generalized ARMA models with martingale difference errors. {\it Journal of Econometrics} \textbf{189}: 492--506.



\end{thebibliography}


\newpage
\begin{table*}\begin{center}
Table 1: ACF values for the simulated data.\\
\label{tab:5}
\footnotesize
\begin{tabular}{*{9}{c}}
\hline \multicolumn{1}{c}{Setting}&\multicolumn{1}{c}{$n$}&\multicolumn{1}{c}{}&
\multicolumn{6}{l}{RRC-GARCH$(p_1,p_2)$ }\\
\cmidrule(l){4-9}
   &  && $M_1$  &  $M_2$     & $M_3$  & $M_4$  &$M_5$  &$M_6$ \\
 \hline
 (a)   &500&$\rho_{X_t}(1)$  &0.392 &0.504  &0.481   &0.221 &0.480 &0.488   \\
               &&$\rho_{X_t}(2)$  &0.180 &0.445  &0.331   &0.406 &0.647 &0.650  \\\\
 (b)   &500&$\rho_{X_t}(1)$  &-0.497 &-0.458  &-0.392   &-0.169   &-0.100    &-0.062   \\
             &&$\rho_{X_t}(2)$  &0.203   &0.333  &0.102   &-0.441  &-0.350    &-0.385       \\
                 \hline
\end{tabular}\end{center}
\end{table*}

\begin{table*}\begin{center}
Table 2: Mean  of estimates, RMSE (within parentheses) for the RRC-GARCH models with Laplace link function and $\tau=0.5$ under the setting (a).\\
\label{tab:1}
\footnotesize
\begin{tabular}{*{10}{c}}
\hline
$M(p_1,p_2)$& n & Method &$c$            & $\phi_1$     & $\phi_2$ & $\psi_1$ & $\psi_2$ & $\tau$ & $\sigma_\zeta^2$\\
\hline
$M_1$(1,0)    &   &  True Value     &-0.4     &0.5   &          & &  &0.5 &0.5\\
     &200&  OLS  &-0.3947 &0.4867  &          & &&  0.3250 & 0.5187 \\
     &   &       &(0.1163)&(0.0955)&      & & &(0.3813) &(0.0864)\\
     &   &  OWLS  &-0.3970   &0.4898  &          && & &\\
     &   &       &(0.1140)&(0.0933)&      && & &\\
     &500&  OLS  &-0.3964  &0.4918  &          & && 0.4242 & 0.5084 \\
     & &         &(0.0731)&(0.0618)&       & && (0.2984)&(0.0556)\\
     &   &  OWLS  &-0.3980  &0.4939  &          && & &\\
     &   &       &(0.0721)&(0.0607)&      && & &\\\\
$M_2$(1,1)   & &   True Value  &-0.4&0.4&          &0.4 & & 0.5 &0.5 \\
     &200&  OLS  &-0.3272  &0.4002  &          &0.3507 &&   0.4600& 0.5186 \\
     &   &       &(0.1607)&(0.0750)&      &(0.1346)& &(0.2163) &(0.1195)\\
     &   &  OWLS  &-0.3346   &0.4020  &          &0.3545& & &\\
     &   &       &(0.1542)&(0.0733)&      &(0.1291)&& &\\
    &500&  OLS  &-0.3740&0.4010  &          &0.3818 && 0.4895 & 0.5063 \\
     & &         &(0.0884)&(0.0471)&       &(0.0792) && (0.1387)&(0.0788)\\
     &   &  OWLS  &-0.3758  &0.4027  &        &0.3813& & &\\
     &   &       &(0.0846)&(0.0444)&      &(0.0756)& & &\\\\
$M_3$(1,2)   & &  True Value     &-0.4 &0.4&     &0.1 &0.4&0.5&0.5 \\
     &200&  OLS  &-0.1106  &0.4147  &          &0.0957 &0.2903&  0.4606 &0.5926 \\
     &   &       &(0.4604)&(0.0772)&      &(0.2325) & (0.2233)&(0.2105) &( 0.2700)\\
     &   &  OWLS  &-0.0872   &0.4219  &          &0.0938&0.2778 & &\\
     &   &       &(0.4753)&(0.0762)&      &(0.2303)&(0.2251) & &\\
     &500&  OLS  &-0.3043  &0.4109  &          &0.0972 &0.3601& 0.4841& 0.5370 \\
     & &         &(0.1639)&(0.0466)&       &(0.1227) &(0.1169)& (0.1311)&(0.1556)\\
     &   &  OWLS  &-0.2933  &0.4158  &          &0.0960&0.3530 & &\\
     &   &       &(0.1711)&(0.0461)&      &(0.1169)&(0.1138) & &\\
  \hline
\end{tabular}\end{center}
\end{table*}

\begin{table*}\begin{center}
Table 2 (continued): Mean  of estimates, RMSE (within parentheses) for the RRC-GARCH models with Laplace link function and $\tau=0.5$ under the setting (a).\\
\label{tab:2c}
\footnotesize
\begin{tabular}{*{10}{c}}
\hline
$M(p_1,p_2)$& n & Method &$c$            & $\phi_1$     & $\phi_2$ & $\psi_1$ & $\psi_2$ & $\tau$ & $\sigma_\zeta^2$\\
\hline
$M_4$(2,0)   & & True Value   &-0.4&0.2&0.5    & & &0.5&0.5\\
     &200&  OLS  &-0.3607  &0.1858  &0.4749          & &&  0.4141 & 0.5165 \\
     &   &       &(0.1263)&(0.0751)&(0.0798)      & & &(0.2678) &(0.1002)\\
     &   &  OWLS  &-0.3700   &0.1890  &0.4800          && & &\\
     &   &       &(0.1255)&(0.0729)& (0.0785)     && & &\\
     &500&  OLS  &-0.3851  &0.1962  &0.4897          & && 0.4629 & 0.5056 \\
     & &         &(0.0809)&(0.0469)&(0.0522)       & && (0.1599)&(0.0619)\\
     &   &  OWLS  &-0.3899  &0.1978  &0.4923          && & &\\
     &   &       &(0.0782)&(0.0453)&(0.0499)      && & &\\\\
$M_5$(2,1)   & & True Value   &-0.4&0.1&0.4    &0.4& &0.5&0.5\\
     &200&  OLS  &-0.2505  &0.0967  & 0.4074   &0.3430 && 0.4911  & 0.5405 \\
     &   &       &(0.2714)&(0.0745)&(0.0843)      &(0.1373) & &(0.1833) &(0.2091)\\
     &   &  OWLS  &-0.2530   &0.0989 &0.4109  &0.3388 & & &\\
     &   &       &(0.2670)&(0.0719)& (0.0806)     &(0.1320)& & &\\
     &500&  OLS  &-0.3450 &0.1003&0.4017         &0.3787 && 0.4891 & 0.5239 \\
     & &         &(0.1211)&(0.0484)& (0.0567)      &(0.0770) && (0.1074)&(0.1267)\\
     &   &  OWLS  &-0.3467  &0.1010  &0.4052      &0.3752& & &\\
     &   &       &(0.1176)&(0.0451)&(0.0529)      &(0.0738)& & &\\\\
$M_6$(2,2)   & & True Value   &-0.4&0.1&0.4    &0.1&0.3&0.5&0.5\\
     &200&  OLS  &-0.1432  &0.1084  &0.4065          &0.0975 &0.1981&  0.4805  &0.5547  \\
     &   &       &(0.4056)&(0.0791)&   (0.0786)   &(0.2231) &(0.2076) &(0.1964) &(0.2314)\\
     &   &  OWLS  &-0.1402   &0.1099  &0.4084        &0.0935&0.1985 & &\\
     &   &       &(0.4094)&(0.0751)&   (0.0760)   &(0.2111)&(0.1996) & &\\
     &500&  OLS  &-0.3140  &0.1018  & 0.4064     &0.1066 &0.2563& 0.4897 &0.5221  \\
     & &         &(0.1620)&(0.0464)&   (0.0467)    & (0.1224)&(0.1151)& (0.1167)&(0.1343)\\
     &   &  OWLS  &-0.3122  &0.1041  &0.4087       &0.1017&0.2564 &&\\
     &   &       &(0.1587)&(0.0437)&  (0.0441)  &(0.1123)&(0.1084) & &\\\\
  \hline
\end{tabular}\end{center}
\end{table*}

\begin{table*}\begin{center}
Table 3: Mean  of estimates, RMSE (within parentheses) for the RRC-GARCH models with Laplace link function and $\tau=0.5$ under the setting (b).\\
\label{tab:3}
\footnotesize
\begin{tabular}{*{10}{c}}
\hline
$M(p_1,p_2)$& n & Method &$c$            & $\phi_1$     & $\phi_2$ & $\psi_1$ & $\psi_2$ & $\tau$ & $\sigma_\zeta^2$\\
\hline
$M_1$(1,0)    &   &  True Value     &2.0     &-0.5   &          & &  &0.5 &0.5\\
    &200&  OLS  &2.0002 &-0.5022  &          & && 0.5109  & 0.5097 \\
     &   &       &(0.1520)&(0.0689)&      && &(0.2361) &(0.1587)\\
     &   &  OWLS  &1.9978 &-0.4999  &          && & &\\
     &   &       &(0.1462)&(0.0633)&      &&& &\\
     &500&  OLS  &2.0007  &-0.5002  &          & &&0.4969  & 0.5114 \\
     & &         &(0.0955)&(0.0425)&       & && (0.1599)&(0.1110)\\
     &   &  OWLS  &2.0005  &-0.4999  &          && & &\\
     &   &       &(0.0920)&(0.0398)&      && & &\\\\
$M_2$(1,1)   & &   True Value  &2.0&-0.4&          &-0.4 & & 0.5 &0.5 \\
     &200&  OLS  &1.9591  & -0.4135 &          &-0.3659 &&  0.4831 & 0.5071 \\
     &   &       &(0.1920)&(0.0826)&      &(0.1334)& &(0.2370) &(0.1304)\\
     &   &  OWLS  &1.9591   &-0.4104  &          &-0.3676& & &\\
     &   &       &(0.1869)&(0.0787)&      &(0.1316)&& &\\
     &500&  OLS  &1.9824  & -0.4072 &          &-0.3835 && 0.5012 & 0.4990 \\
     & &         &(0.1110)&(0.0490)&       &(0.0790) && (0.1623)&(0.0866)\\
     &   &  OWLS  &1.9837  &-0.4055  &          &-0.3852& & &\\
     &   &       &(0.1078)&(0.0467)&      &(0.0765)& & &\\\\
$M_3$(1,2)   & &  True Value     &2.0 &-0.4&     &-0.1 &-0.4&0.5&0.5 \\
     &200&  OLS  &1.8010  &-0.4200  &          &-0.0365 &-0.3126& 0.4677  &0.5041  \\
     &   &       &(0.4313)&(0.0793)&      &(0.1736) & (0.1921)&(0.2598) &(0.1194)\\
     &   &  OWLS  &1.8745   &-0.4141  &          &-0.0617&-0.3421 & &\\
     &   &       &(0.4210)&(0.0755)&      &(0.1810)&(0.1804) & &\\
     &500&  OLS  &1.9454  &-0.4057  &          &-0.0814 &-0.3777&  0.4840& 0.5040 \\
     & &         &(0.2451)&(0.0480)&       &(0.1031) &(0.1025)& (0.1863)&(0.0873)\\
     &   &  OWLS  &1.9545  &-0.4043  &          & -0.0842&-0.3817 & &\\
     &   &       &(0.2378)&(0.0461)&      &(0.1022)&(0.0988) & &\\
  \hline
\end{tabular}\end{center}
\end{table*}

\begin{table*}\begin{center}
Table 3 (continued): Mean  of estimates, RMSE (within parentheses) for the RRC-GARCH models with Laplace link function and $\tau=0.5$ under the setting (b).\\
\label{tab:3c}
\footnotesize
\begin{tabular}{*{10}{c}}
\hline
$M(p_1,p_2)$& n & Method &$c$            & $\phi_1$     & $\phi_2$ & $\psi_1$ & $\psi_2$ & $\tau$ & $\sigma_\zeta^2$\\
\hline
$M_4$(2,0)   & & True Value   &2.0&-0.2&-0.5    & & &0.5&0.5\\
     &200&  OLS  &2.0143  &-0.2021  &-0.5066          & && 0.4961  & 0.5018 \\
     &   &       &(0.1955)&(0.0662)&(0.0723)      & & &(0.2500) &(0.1413)\\
     &   &  OWLS  &2.0115   &-0.2021  &-0.5040          && & &\\
     &   &       &(0.1881)&(0.0635)& (0.0677)     && & &\\
     &500&  OLS  &2.0031  &-0.2003  & -0.5030         & && 0.4965 & 0.5037 \\
     & &         &(0.1216)&(0.0400)&(0.0458)       & && (0.1650)&(0.0935)\\
     &   &  OWLS  &1.9999  &-0.1994  &-0.5016          && & &\\
     &   &       &(0.1162)&(0.0382)&(0.0430)      && & &\\\\
$M_5$(2,1)   & & True Value   &2.0&-0.1&-0.4    &-0.4& &0.5&0.5\\
     &200&  OLS  & 1.9343 & -0.1122 & -0.4074         &-0.3395 && 0.4607  & 0.5121 \\
     &   &       &(0.3210)&(0.0735)&(0.0725)      &(0.1987) & &(0.2744) &(0.1295)\\
     &   &  OWLS  &1.9488   &-0.1112  & -0.4059         &-0.3503&& &\\
     &   &       &(0.3171)&(0.0720)& (0.0704)     &(0.1958)& & &\\
     &500&  OLS  &1.9797  &-0.1027  &-0.4021          &-0.3830 &&  0.4903& 0.5010 \\
     & &         &(0.1733)&(0.0436)& (0.0440)      &(0.1035) && (0.1860)&(0.0838)\\
     &   &  OWLS  &1.9806  &-0.1021  &-0.4010          &-0.3846& & &\\
     &   &       &(0.1683)&(0.0428)&(0.0424)      &(0.1007)& & &\\\\
$M_6$(2,2)   & & True Value   &2.0&-0.1&-0.4    &-0.1&-0.3&0.5&0.5\\
     &200&  OLS  &1.9230  &-0.1077  & -0.4117         &-0.0711 &-0.2628& 0.4685 & 0.5040 \\
     &   &       &(0.3051)&(0.0761)&   (0.0886)   &(0.1579) &(0.1766) &(0.2479) &(0.1225)\\
     &   &  OWLS  &1.9318   &-0.1074  &-0.4056          &-0.0725&-0.2714 & &\\
     &   &       &(0.3047)&(0.0747)&   (0.0873)   &(0.1598)&(0.1764) & &\\
     &500&  OLS  &1.9749  & -0.1048 &  -0.4051        &-0.0885 &-0.2855& 0.4833 & 0.5020 \\
     & &         &(0.1705)&(0.0462)&   (0.0512)    & (0.0964)&(0.0962)& (0.1649)&(0.0796)\\
     &   &  OWLS  &1.9763  &-0.1032  &-0.4039          &-0.0910&-0.2863 & &\\
     &   &       &(0.1645)&(0.0441)&  (0.0498)  &(0.0925)&(0.0951) & &\\\\
  \hline
\end{tabular}\end{center}
\end{table*}

\begin{table*}\begin{center}
Table 4 : Frequency of orders selected by AIC and BIC for the RRC-GARCH models with Laplace link function and $\tau=0.5$ in 1000 realizations under the setting (a).\\
\label{tab:4}
\footnotesize
\begin{tabular}{*{9}{c}}
\hline \multicolumn{1}{c}{Model}&\multicolumn{1}{c}{ n
}&\multicolumn{1}{c}{ }&
\multicolumn{5}{l}{$(p_1,p_2)$ }\\
\cmidrule(l){4-9}
 &  &   & (1,0)  &  (1,1)     & (1,2)  & (2,0)  &(2,1)  &(2,2) \\
 \hline $M_1$   &100 & AIC  &453   &128    &130   &105    &93    &91      \\
                &    & BIC  &755   &75    &64   &52    &37   &17     \\
        &200 & AIC  &490  &115   &129   &89   &105   &72   \\
        &    & BIC  &803  &55   &46   &46   &37  &13  \\
        &500 & AIC  &545  &129  &101  &66   &104   &55   \\
        &    & BIC  &878  &46   &13   &29  &30   &4   \\
        \\
  \hline $M_2$   &100 & AIC  &246   &301    &113   &197    &57    &86      \\
                &    & BIC  &504   &214     &45    &193    &22   & 22     \\
        &200 & AIC  &78  &444   &116   &247   &37   &78   \\
        &    & BIC  &254  &423   &35   &268   &8  &12  \\
        &500 & AIC  &3  &562  &108  &163   &60   &104   \\
        &    & BIC  &25 &698  &24   &237  &8   &8   \\
        \\
  \hline $M_3$   &100 & AIC  &224   &263    &271     &82  &64   &96      \\
                 &    & BIC  &485   &273    &110     &75   &26  &31     \\
        &200 & AIC  &46  &282   &454   &23   &65   &130   \\
        &    & BIC  &190  &437   &279   &39   &36  &19  \\
        &500 & AIC  &0  &65  &683  &0   &43   &209   \\
        &    & BIC  &9  &225 &703  &1   &43   &19   \\
        \\
 \hline $M_4$   &100 & AIC  &6   &29    &47   &581    &159    &178      \\
                &    & BIC  &24   &32     &28  &785   &65   &66     \\
        &200 & AIC  &1  &4   &20  &635   &151   &189   \\
        &    & BIC  &3  &5   &13   &890   &49  &40  \\
        &500 & AIC  &0  &0  &2  &656   &179   &163   \\
        &    & BIC  &0  &1  &1  &933   &44   &21   \\
        \\
  \hline $M_5$   &100 & AIC  &0   &49    &60   &356    &368    &167      \\
                &    & BIC  &7   &79     &49    &591   &222   &52     \\
        &200 & AIC  &0  &5   &50   &98   &670   &177   \\
        &    & BIC  &0  &28  &48   &308   &573  &43  \\
        &500 & AIC  &0  &0  &7  &6   &789   &198   \\
        &    & BIC  &0  &2  &6  &19   &943   &30   \\
        \\
  \hline $M_6$   &100 & AIC  &2   &45    &41   &445    &261    &206      \\
                &    & BIC  &13   &78    &33   &686   &151   &39     \\
        &200 & AIC  &0  &15   &20   &176   &378   &411   \\
        &    & BIC  &0  &41   &15   &484   &340  &120  \\
        &500 & AIC  &0  &0  &4  &7   &155   &834  \\
        &    & BIC  &0  &1  &4  &84  &405   &506   \\
        \hline
\end{tabular}\end{center}
\end{table*}

\begin{table*}\begin{center}
Table 5 : Frequency of orders selected by AIC and BIC for the RRC-GARCH models with Laplace link function and $\tau=0.5$ in 1000 realizations under the setting (b).\\
\label{tab:5}
\footnotesize
\begin{tabular}{*{9}{c}}
\hline \multicolumn{1}{c}{Model}&\multicolumn{1}{c}{ n
}&\multicolumn{1}{c}{ }&
\multicolumn{5}{l}{$(p_1,p_2)$ }\\
\cmidrule(l){4-9}
 &  &   & (1,0)  &  (1,1)     & (1,2)  & (2,0)  &(2,1)  &(2,2) \\
 \hline $M_1$   &100 & AIC  &578   &117    &103   &122    &32    &48      \\
                &    & BIC  &852   &45     &32   &63     &4     &4     \\
        &200 & AIC  &640  &93   &81   &116   &42   &28   \\
        &    & BIC  &903  &27   &15   &49   &5  &1 \\
        &500 & AIC  &591  &110  &110  &133   &34   &22   \\
        &    & BIC  &932  &26   &7   &35  &0   &0   \\

        \\
  \hline $M_2$   &100 & AIC  &289   &290    &185   &171    &35    &30      \\
                &    & BIC  &552   &221     &61    &152   &9   &5     \\
        &200 & AIC  &86  &446   &196   &182   &57   &33   \\
        &    & BIC  &291  &440   &62   &183   &19  &5  \\
        &500 & AIC  &3  &564  &202  &123   &70   &38   \\
        &    & BIC  &34  &739 &49   &154  &22   &2   \\
        \\
  \hline $M_3$   &100 & AIC  &433   &54    &223   &122    & 30   &138      \\
                &    & BIC  &761   &16     &99     &87   &8   &29     \\
        &200 & AIC  &295 &7   &459   &74   &23   &142   \\
        &    & BIC  &673  &2   &244   &43   &9  &29  \\
        &500 & AIC  &56  &1  &753  &11   &15   &164   \\
        &    & BIC  &300  &0   &648   &17  &8   &27   \\
        \\
 \hline $M_4$   &100 & AIC  &0   &11    &39   &616    &184    &150      \\
                &    & BIC  &1   &13    &23   &854    &78   &31    \\
        &200 & AIC  &0  &0   &2   &688   &160   &150   \\
        &    & BIC  &0  &0   &2   &929   &50  &19  \\
        &500 & AIC  &0  &0  &0  &729   &158   &113   \\
        &    & BIC  &0  &0  &0  &965   &30   &5   \\
        \\
 \hline $M_5$   &100 & AIC  &4   &15    &46   &417    &335    &183      \\
                &    & BIC  &22   &22   &35   &672   &204   &45    \\
        &200 & AIC  &0  &3   &11   &247   &572   &167   \\
        &    & BIC  &1  &2   &7   &542   &410  &38  \\
        &500 & AIC  &0  &0  &24  &48   &734   &194   \\
        &    & BIC  &0  &0  &24  &220  &729   &27   \\
        \\
 \hline $M_6$   &100 & AIC  &2   &7    &41   &577    &95    &278      \\
                &    & BIC  &5   &9    &31     &809   &39   &107    \\
        &200 & AIC  &0  &1   &9   &407   &92   &491   \\
        &    & BIC  &0  &1   &8   &769   &39  &183  \\
        &500 & AIC  &0  &0  &0  &99   &39   &862   \\
        &    & BIC  &0  &0  &1  &474  &26   &499   \\
        \hline
\end{tabular}\end{center}
\end{table*}

\begin{table*}\begin{center}
Table 6: AIC and BIC values for the real data.\\
\label{tab:6}
\footnotesize
\begin{tabular}{*{9}{c}}
\hline \multicolumn{1}{c}{Data}&\multicolumn{1}{c}{$n$}&\multicolumn{1}{c}{}&
\multicolumn{6}{l}{RRC-GARCH$(p_1,p_2)$ }\\
\cmidrule(l){4-9}
   &  && (1,0)  &  (1,1)     & (1,2)  & (2,0)  &(2,1)  &(2,2) \\
 \hline
Ecoli counts   &616&AIC  &2369.850 &2289.293  & 2283.812  & \textbf{2280.248}&2283.892 &2283.049   \\
               &&BIC  &2387.523 &2311.385  &2310.312   &\textbf{2302.331} &2310.393 &2313.966  \\\\
WPP counts   &2825&AIC  &9088.451   &8880.538  &8833.220   &8937.021   &\textbf{8811.859}    &8818.571   \\
             &&BIC  &9112.231   &8910.264  &8868.889   &8966.746   &\textbf{8847.528}    &8860.185       \\
                 \hline
\end{tabular}\end{center}
\end{table*}

\begin{table*}\begin{center}
Table 7: Estimates and their estimated standard deviations (in parentheses) for the real data.\\
\label{tab:7}
\footnotesize
\begin{tabular}{*{10}{c}}
\hline
Data& $n$ &Model&Estimate&$c$ & $\phi_1$ &$\phi_2$ & $\psi_1$ & $\tau$ & $\sigma_\varsigma^2$\\
\hline
Ecoli counts&  616  &(2,0) & OLS &4.8473 &0.4833 &0.2468 &  & 0.9999 &0.1039\\
            &    & &  &(1.3041) &(0.0796) &(0.0692) & &   (0.0955)&(0.0688)\\
            &    &     & OWLS &6.5702 &0.3983 &0.2442 & &   &\\
            &    & &  &(0.8509) &(0.0588) &(0.0480) & &   &\\
WPP counts&  2825  &(2,1)  &OLS &-0.1794 &0.3162 &-0.1271 &0.7478  &0.6845&1.3986\\
          &    &  & &(0.1077) &(0.0253) &(0.0333) &(0.0335)  &  (0.0796)&(0.4956)\\
          &    &   &OWLS &-0.1501 &0.3111 &-0.1069 &0.7294  &  &\\
          &    &  & &(0.1061) &(0.0212) &(0.0301) &(0.0331)  &  &\\
  \hline
\end{tabular}\end{center}
\end{table*}

\begin{table*}\begin{center}
Table 8:  Model diagnostics of the real data. Sample mean of the standardized residuals: $E_n(r_t)$; Sample SD of the standardized residuals: $Sd_n(r_t)$; the maximum absolute value of the sample autocorrelation: $\max_{1\leq k\leq 20}|\rho_{r_t}(k)|$.\\
\label{tab:8}
\footnotesize
\begin{tabular}{*{9}{c}}
\hline
Data& $n$ &Model&Method&$E_n(r_t)$&$Sd_n(r_t)$ &$\max_{1\leq k\leq 20}|\rho_{r_t}(k)|$& MAR & MSPR\\
\hline
Ecoli counts&  616  &(2,0) &OLS&0.0222  &1.0824 &0.134 &5.2357 & 1.1702 \\
            &    &         &OWLS&0.0034  &1.0593 &0.097 &5.2357 &1.1203  \\
WPP counts  &  2825 &(2,1)&OLS&0.0007 &1.0345 &0.062 &3.9003 & 1.0697\\
            &    &         &OWLS&0.0005 &1.0348 &0.062 &3.9003 &1.0704  \\
  \hline
\end{tabular}\end{center}
\end{table*}

\begin{table*}\begin{center}
Table 9: Predictions of the real data.\\
\label{tab:9}
\footnotesize
\begin{tabular}{*{9}{c}}
\hline
Data& $n_{new}$ &Model&Method&$E_{n_{new}}(r_t)$&$Sd_{n_{new}}(r_t)$ &$\max_{1\leq k\leq 14}|\rho_{r_t}(k)|$& MAR & MSPR\\
\hline
Ecoli counts&  30  &(2,0) &OLS&-0.1000 &1.1125   &0.441&5.3518 &1.2065  \\
            &      &      &OWLS &-0.1434   &1.0623  &0.426 &5.2252 &1.1114  \\
WPP counts  &  100  &(2,1)&OLS&0.0179    &1.0962  &0.320 &3.5827 &1.1899  \\
            &    &        &OWLS&0.0165    &1.0939 &0.322  &3.5785 &1.1848  \\
  \hline
\end{tabular}\end{center}
\end{table*}

\begin{center}
\begin{figure}
\centerline{\includegraphics[width=0.8\textwidth, angle=
0]{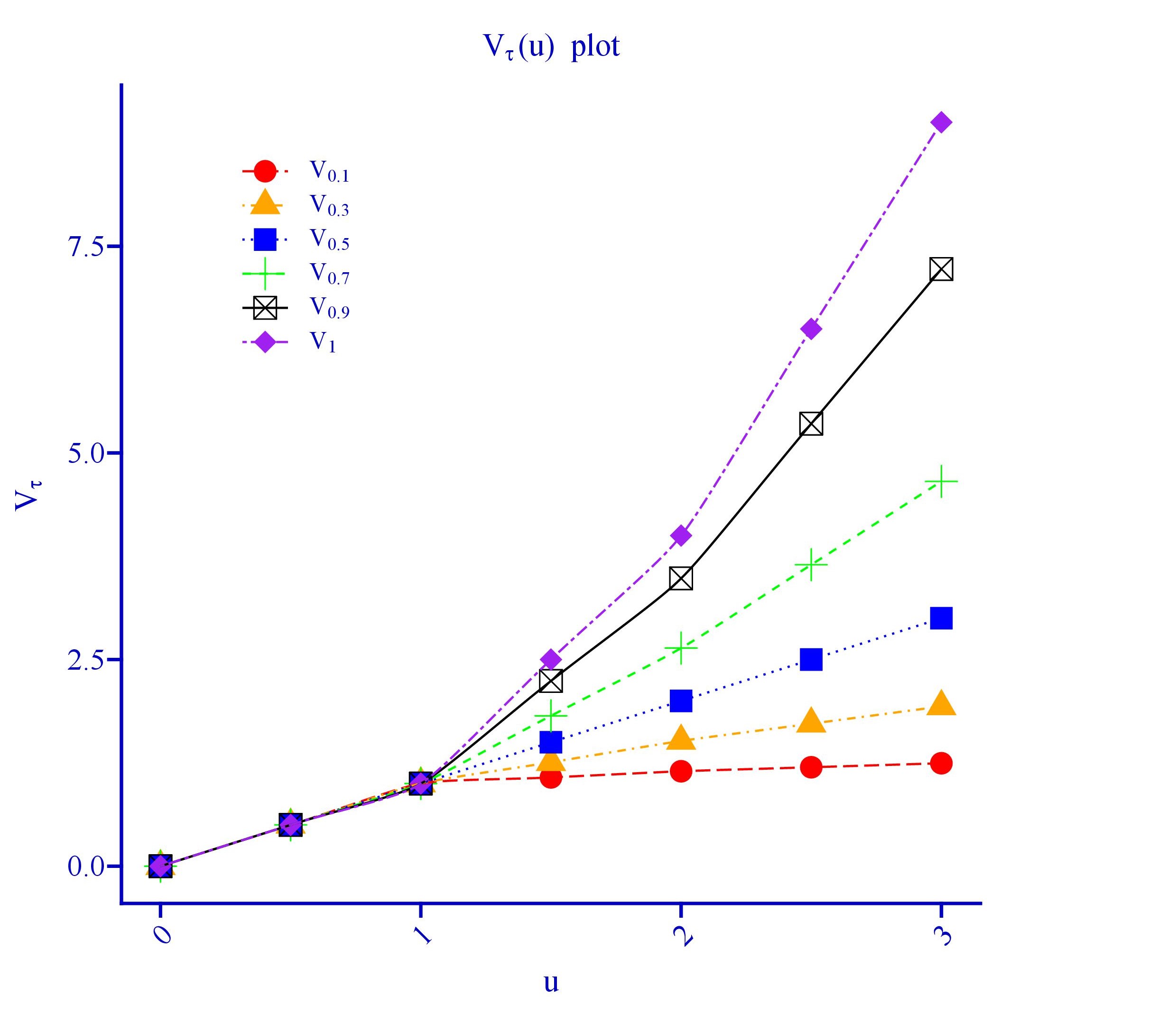}} \caption{$L_\sigma(u)$ plots with $u\in[-2,2]$ and $\sigma=0,0.2,0.4,0.6,0.8$ and $1.0$.}\label{Lplot}
\end{figure}
\end{center}

\begin{center}
\begin{figure}
\centerline{\includegraphics[width=0.8\textwidth, angle=
0]{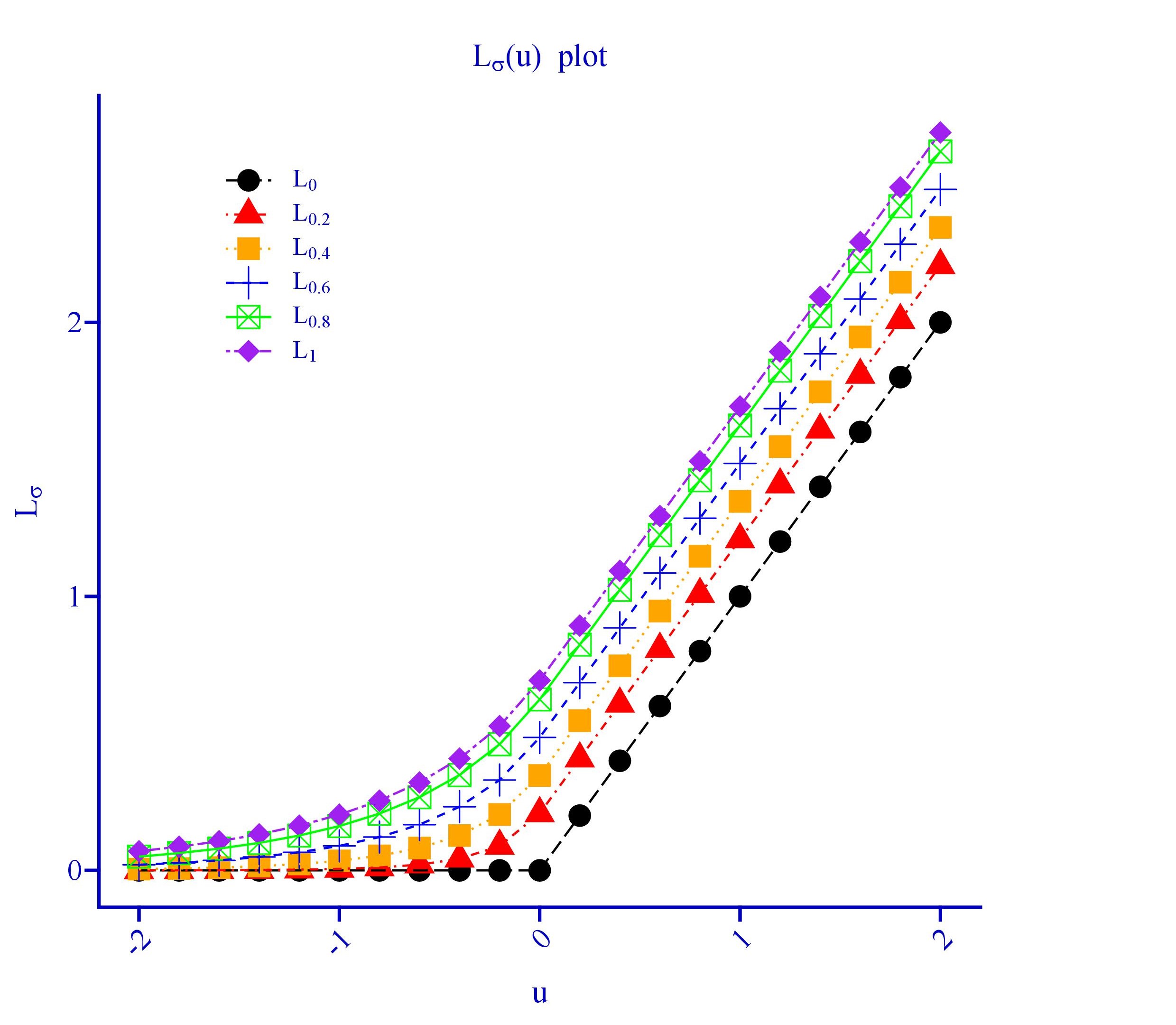}} \caption{$L_\sigma(u)$ plots with $u\in[-2,2]$ and $\sigma=0,0.2,0.4,0.6,0.8$ and $1.0$.}\label{Lplot}
\end{figure}
\end{center}

\begin{center}
\begin{figure}
\centerline{\includegraphics[width=0.8\textwidth, angle=
0]{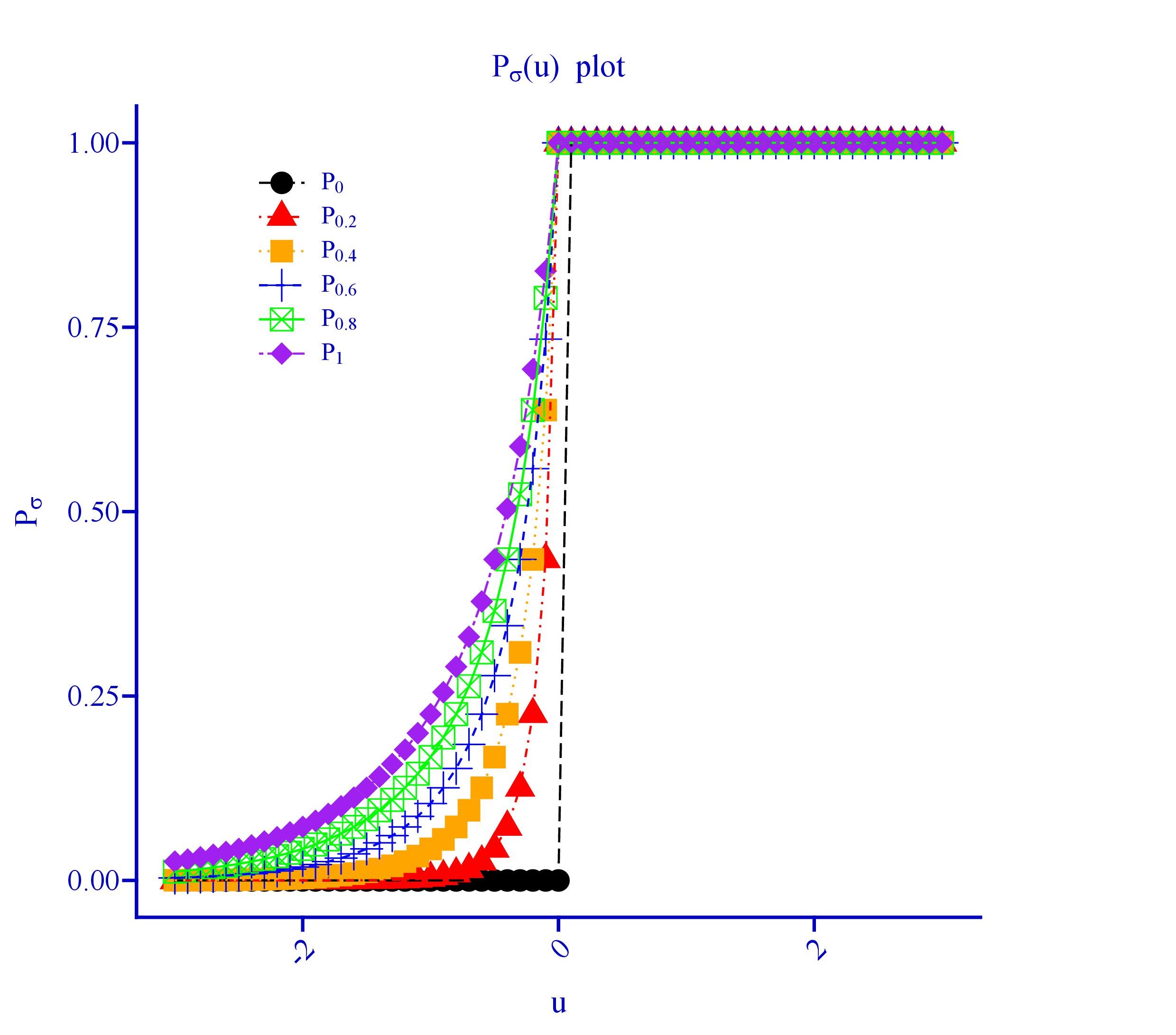}} \caption{$P_\sigma(u)$ plots with $u\in[-2,2]$ and $\sigma=0,0.2,0.4,0.6,0.8$ and $1.0$.}\label{Dplot}
\end{figure}
\end{center}

\begin{center}
\begin{figure}
\centerline{\includegraphics[width=1.0\textwidth, angle=
0]{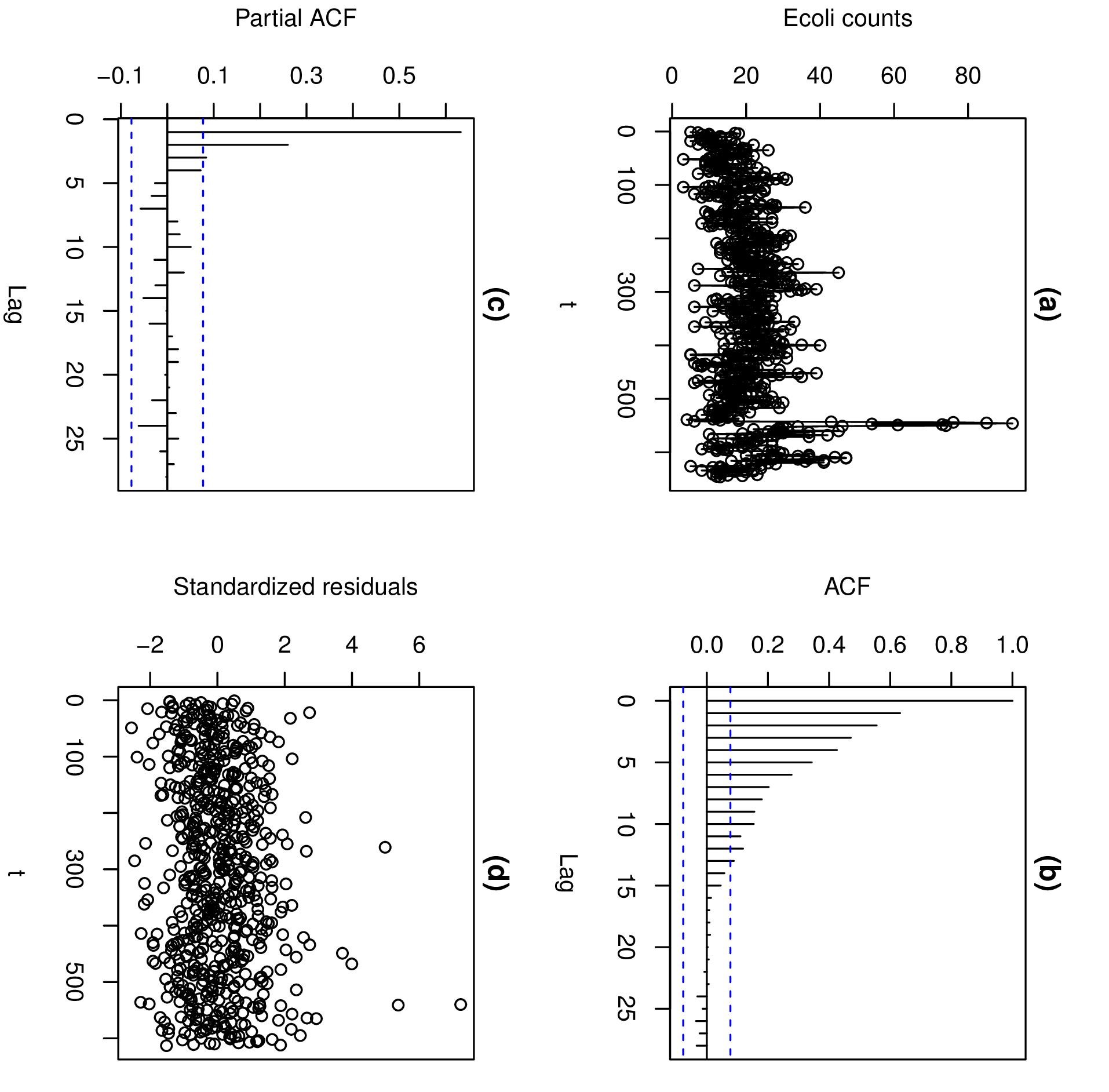}} \caption{Ecoli counts from Section 7.1: (a) time series plot; (b) sample ACF against Lag; (c) sample PACF against Lag; and (d) standardized residuals plot using the OLS estimates of $\vtheta$ and $\vlambda$.}\label{Ecoli}
\end{figure}
\end{center}

\begin{center}
\begin{figure}
\centerline{\includegraphics[width=1.0\textwidth, angle=
0]{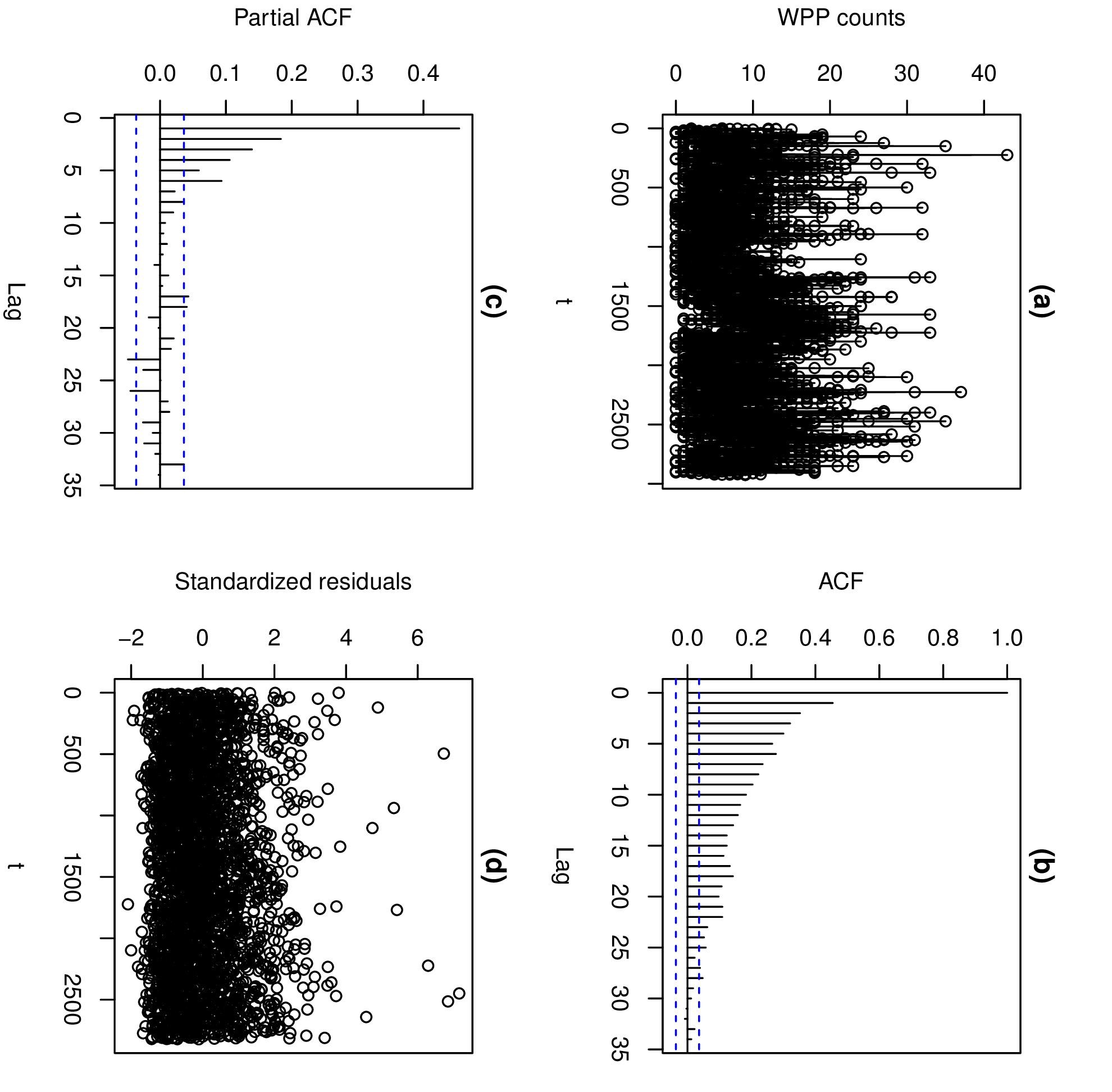}} \caption{WPP counts from Section 7.2: (a) time series plot; (b) sample ACF against Lag; (c) sample PACF against Lag; and (d) standardized residuals plot using the OLS estimates of $\vtheta$ and $\vlambda$.}\label{WPP}
\end{figure}
\end{center}

\end{document}